\documentclass[aps,showpacs,nofootinbib,superscriptaddress]{revtex4}
\usepackage{graphicx}
\usepackage{dcolumn}

 \def\tstrut{\vrule height2.5ex depth0pt width0pt} % used in tables
\begin{document}

\title{ Charmed and Bottom Baryons: a Variational Approach based on Heavy
Quark Symmetry. } \author{C. Albertus}
\affiliation{Departamento de F\'{\i}sica Moderna, Universidad de
Granada, E-18071 Granada, Spain.}  \author {J.E. Amaro}
\affiliation{Departamento de F\'{\i}sica Moderna, Universidad de
Granada, E-18071 Granada, Spain.}  \author{ E. Hern\'andez}
\affiliation{Grupo de F\'\i sica Nuclear, Facultad de Ciencias,
E-37008 Salamanca, Spain.}  \author {J. Nieves}
\affiliation{Departamento de F\'{\i}sica Moderna, Universidad de
Granada, E-18071 Granada, Spain.}
\begin{abstract} 
\rule{0ex}{3ex} The use of Heavy Quark Symmetry to study bottom and
charmed baryons leads to important simplifications of the
non-relativistic three body problem, which turns out to be easily
solved by a simple variational ansatz.  Our simple scheme reproduces
previous results (baryon masses, charge and mass radii,
%, charge form factors
$\cdots$) obtained by solving the Faddeev
equations with simple non-relativistic quark--quark potentials,
adjusted to the light and heavy--light meson spectra. Wave functions,
parameterized in a simple manner, are also given and thus they can be
easily used to compute further observables.  Our method has been also used to 
find
%,
%for the very first time,
 the predictions for strangeness-less baryons
of the SU(2) chirally inspired quark-quark interactions. We 
%compare
%these results with those obtained from phenomenological potentials and
find that the one pion exchange term of the chirally inspired
interactions leads to relative changes of the $\Lambda_b$ and
$\Lambda_c$ binding energies as large as 90\%.

\end{abstract}

\maketitle

\section{Introduction}
%************************

Prior, and at around, the time the charmed baryons existence was confirmed
\cite{Baltay79} a great theoretical activity developed in  order to understand
and predict their masses~\cite{CIK79,MI80,RT83}, magnetic 
moments~\cite{CJ76}, or decay 
properties~\cite{VZO75,Buras76,KKW79}. This activity continued in the following
decades.
The discovery of the $\Lambda_b$ baryon at CERN~\cite{Al91}, the
discovery of most of the charmed baryons of the SU(3) multiplet on the
second level of the SU(4) lowest 20-plet~\cite{pdg02}, and the claims
of indirect evidences for the semileptonic decays of $\Lambda_b$ and
$\Xi_b$~\cite{semi} renewed the interest
%led to an increased 
in the spectroscopy
and weak decays of heavy baryons.  There are eight lowest-lying
baryons containing one heavy and two light quarks (up, down, or
strange). The quantum numbers of the charmed and bottom baryons are
listed in Table~\ref{tab:summ}.  

Heavy Quark Symmetry (HQS) has proved
to be a useful tool to understand the bottom and charm
physics~\cite{IW89,Ne94,KKP94}, and it has been extensively used to describe
the dynamics of systems containing a heavy quark $c$ or $b$. For instance, all
lattice QCD simulations rely on HQS to
describe bottom systems~\cite{UKQCD}. 
%However, HQS has not been
%systematically employed in the context of non-relativistic quark
%models. 
HQS is an approximate SU($N_F$) symmetry of QCD
, being $N_F$
the number of heavy flavors: $c$, $b$, $\cdots$, which appear in
systems containing heavy quarks with masses that are much larger
than the typical quantities ($q=\Lambda_{QCD}$, $m_u,m_d,m_s\cdots$ )
which set up the energy scale of the dynamics of the remaining degrees
of freedom. HQS has some resemblances, in atomic physics, to
the approximate independence of the electron properties from the
nuclear spin and mass, for a fixed nuclear charge.
Up to corrections of the order\footnote{The quantities $q$ and $m_Q$
are a typical energy scale relevant for the light degrees of freedom
and the mass of the heavy quark, respectively.} ${\cal O}(q/m_Q)$, HQS
guaranties that the heavy baryon light degrees of freedom quantum
numbers, compiled in Table~\ref{tab:summ}, are always well
defined. The symmetry also predicts that the pair of baryons $\Sigma$
and $\Sigma^*$ or the pair $\Xi^\prime$ and $\Xi^*$ or the pair
$\Omega$ and $\Omega^*$ become degenerated for an infinitely massive
heavy quark, since both baryons have the same cloud of light degrees
of freedom.
\begin{table}

\begin{tabular}{cccccccc}\hline
Baryon &~~~~$S$~~~~&~~~~$J^P$~~~~&~~~~ $I$~~~~&~~~~$S_{\rm light}^\pi$~~~~& 
Quark content 
& $M_{exp.}$ \cite{pdg02}~~~& $M_{Latt.}$~\cite{bowler}
\\
       &       &         &   &          &               & [MeV]  & [MeV]   
\\\hline
$\Lambda_c$& 0 &$\frac12^+$& 0 &$0^+$&$udc$& $2285 \pm 1$ & $2270 \pm 50  $
\\
$\Sigma_c$ & 0 &$\frac12^+$& 1 &$1^+$&$llc$& $2452 \pm 1$ & $2460 \pm 80 $
\\
$\Sigma^*_c$ & 0 &$\frac32^+$& 1 &$1^+$&$llc$& $2518 \pm 2$ &$ 2440
\pm 70 $
\\
$\Xi_c$ & $-$1 &$\frac12^+$&$\frac12$&$0^+$&$lsc$& $2469 \pm 3$ & $2410
\pm 50 $
\\
$\Xi'_c$ & $-$1 &$\frac12^+$&$\frac12$&$1^+$&$lsc$& $2576 \pm 3$ & $2570
\pm 80$ 
\\
$\Xi^*_c$ &$-$1&$\frac32^+$&$\frac12$&$1^+$&$lsc$& $2646 \pm 2$ & $2550
\pm 80$
\\
$\Omega_c$ &$-$2 &$\frac12^+$& 0 &$1^+$&$ssc$& $2698 \pm 3$ & $2680 \pm 70$
\\
$\Omega^*_c$ &$-$2 &$\frac32^+$& 0 &$1^+$&$ssc$&     & $2660 \pm 80$
\\\hline
$\Lambda_b$& 0 &$\frac12^+$& 0 &$0^+$&$udb$& $ 5624 \pm 9$  & $5640
\pm 60 $
\\
$\Sigma_b$ & 0 &$\frac12^+$& 1 &$1^+$&$llb$&  & $5770 \pm 70 $
\\
$\Sigma^*_b$ & 0 &$\frac32^+$& 1 &$1^+$&$llb$&  & $5780 \pm 70$
\\
$\Xi_b$ & $-$1 &$\frac12^+$&$\frac12$&$0^+$&$lsb$&    & $5760 \pm 60 $
\\
$\Xi'_b$ & $-$1 &$\frac12^+$&$\frac12$&$1^+$&$lsb$&    & $5900 \pm 70$
\\
$\Xi^*_b$ &$-$1&$\frac32^+$&$\frac12$&$1^+$&$lsb$&  & $5900 \pm 80 $
\\
$\Omega_b$ &$-$2 &$\frac12^+$& 0 &$1^+$&$ssb$&  & $5990 \pm 70 $
\\
$\Omega^*_b$ &$-$2 &$\frac32^+$& 0 &$1^+$&$ssb$&     & $6000 \pm 70$
\\\hline
\end{tabular}
\caption{Summary of the quantum numbers, experimental and lattice QCD
masses of the baryons containing a single heavy quark. $I$, and
$S_{\rm light}^\pi$ are the isospin, and the spin parity of the light
degrees of freedom and $S$, $J^P$ are strangeness and the spin parity
of the baryon ($l$ denotes a light quark of flavor $u$ or
$d$). Experimental masses are taken from Ref.~\protect\cite{pdg02} and
the isospin averaged masses are quoted, with errors counting for the
mass differences between the members of the same isomultiplet. Errors
on the lattice QCD masses have been obtained by adding in quadratures
the statistical and systematic errors given in
Ref.~\protect\cite{bowler}.}
\label{tab:summ}
\end{table}

%=========================================

However, HQS has not been systematically employed in the context of
non-relativistic quark models.  Non-relativistic quark models, based
upon simple quark-quark potentials, partially inspired by Quantum
Chromodynamics (QCD), lead to reasonably good descriptions of hadrons
as bound states of constituent quarks and also of the general features
of the baryon-baryon interaction~\cite{pro02}--\cite{Sh89}. Most of
the quark-quark interactions include a term with a shape and a color
structure determined from the One Gluon Exchange (OGE)
contribution~\cite{Ru75} and a confinement potential. The force which
confines the quarks is still not well understood, although it is
assumed to come from long-range nonperturbative features of
QCD~\cite{Su95}. Those terms do not incorporate another important
feature of QCD: Chiral Symmetry (CS) and its pattern of spontaneous
breaking. Quark--quark interaction terms derived from Spontaneous
Chiral Symmetry Breaking (SCSB) have been taken into account for the
description of the nucleon--nucleon system and/or the light baryon
spectra~\cite{Fe93,VB94,VG96,GR96,GP98}

In this work, we develop a variational approach for the solution of
the non-relativistic three-body problem in baryons with a heavy
quark. Thanks to HQS, the method we propose turns out to be quite
simple, and leads to simple and manageable wave functions\footnote{We
use a family of wave functions for which the light degrees of freedom
quantum numbers are well defined.}.  We consider several simple
phenomenological quark--quark interactions ~\cite{BD81,Si96} which
free parameters have been adjusted in the meson sector and are then
free of three-body ambiguities. With those potentials we compute the
ground states of some heavy flavor hyperons and show how our simple
approach reproduces the results obtained in Ref.~\cite{Si96} by
solving involved Faddeev type equations. Besides, we also study the
$\Sigma^*$, $\Xi^\prime$, $\Xi^*$ and $\Omega^*$ baryon ground states,
which were not considered in the work of Ref.~\cite{Si96}. After that,
and within our variational framework, we work out the predictions for
the strangeness-less charmed and bottom baryons of the SU(2) chirally
inspired quark-quark interaction of Ref.~\cite{Fe93}, applied with
great success to the meson sector in Ref.~\cite{BFV99}, and study the
effects in these systems of including a pattern of SCSB. Some
preliminary results have been presented in Ref.~\cite{Al03}.

The paper is organized as follows. After this introduction, we study
the hamiltonian of the system, different inter-quark interactions
(Sect.~\ref{sec:3bp}) and the general structure of the variational
baryon wave--function (Sect.~\ref{vwf}). Some static properties (mass
and charge densities, $\cdots$) are revisited in the
Sect.~\ref{sec:sp} and finally in the Sects.~\ref{sec:res} and
\ref{sec:concl} we present our results and compile the main
conclusions of this work, respectively.

\section{Three Body Problem}\label{sec:3bp}
\subsection{Intrinsic Hamiltonian}
\begin{figure}[b]
\vspace{-3cm}
\centerline{\includegraphics[height=25cm]{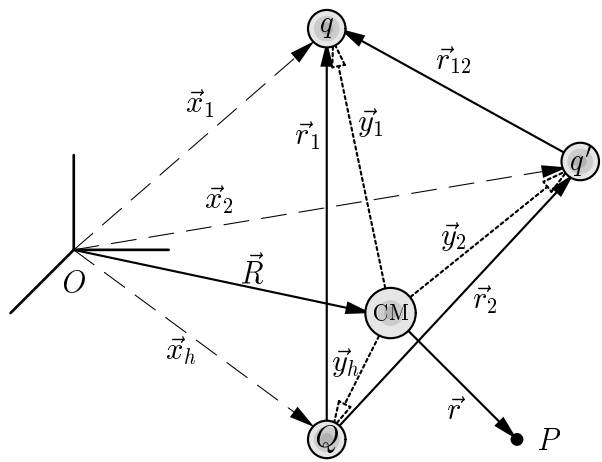}}
\vspace{-15cm}
\caption{ Definition of different coordinates used through this
work.}\label{fig:coor}
\end{figure}

In the Laboratory (LAB) frame (see Fig.~\ref{fig:coor}), the
Hamiltonian ($H$) of the three quark ($q,q^\prime, Q$, with $q,
q^\prime =l$ or $s$ and $Q=c$ or $b$) system reads:
\begin{eqnarray}
H&=&  \sum_{i=q,q^\prime,Q} \left (m_i
-\frac{\vec{\nabla}_{x_i}^2}{2m_i}\right ) +
V_{qq^\prime} + V_{Qq}+V_{Q q^\prime }
\end{eqnarray}
where $m_q,m_{q^\prime}$ and $m_Q$ are the quark masses, and the
quark-quark interaction terms, $V_{ij}$, depend on the quark
spin-flavor quantum numbers and the quark coordinates ($\vec{x}_1,
\vec{x}_2$ and $\vec{x}_h$ for the $q,q^\prime$ and $Q$ quarks
respectively). The nabla operators in the kinetic energy stand for
derivatives with respect to the spatial variables $\vec{x}_1, \vec{x}_2$
and $\vec{x}_h$. To separate the Center of Mass (CM) free motion,
we go to the heavy quark frame ($\vec{R},\vec{r}_1,\vec{r}_2$),
\begin{eqnarray}
\vec{R}&=&\frac{m_q\vec{x}_1 + m_{q^\prime}\vec{x}_2 + m_Q \vec{x}_h}
{m_q+m_{q^\prime}+m_Q} \nonumber \\
\vec{r}_1 & = & \vec{x}_1 - \vec{x}_h \nonumber \\
\vec{r}_2 & = & \vec{x}_2 - \vec{x}_h 
\end{eqnarray}
where $\vec{R}$ and $\vec{r}_1$ ($\vec{r}_2$) are the CM position in
the LAB frame and the relative position of the quark $q$ ($q^\prime$)
with respect to the heavy quark $Q$.   The hamiltonian now reads
\begin{eqnarray}
H&=&
-\frac{\vec\nabla_{\vec{R}}^2}{2 M} +
H^{\rm int} \\ H^{\rm
int}&=&-\frac{\vec\nabla_1^2}{2\mu_1}-\frac{\vec\nabla_2^2}{2\mu_2}-
\frac{\vec\nabla_1\cdot\vec\nabla_2}{m_Q}+
V_{qq^\prime}(\vec{r}_1-\vec{r}_2,spin)+\nonumber\\ &&+V_{Qq}(\vec
r_1,spin)+V_{Qq^\prime}(\vec r_2,spin)+ \sum_{i=q,q^\prime,Q} m_i
\end{eqnarray}
where $M = \left(m_q+m_{q^\prime}+m_Q\right)$, $\mu_{1,2} = \left (
1/m_{q,q^\prime} + 1/m_Q\right)^{-1}$ and $\vec\nabla_{1,2} =
\partial/\partial_{\vec{r}_1,\vec{r}_2}$. The intrinsic hamiltonian
$H^{\rm int}$ describes the dynamics of the baryon and we will use a
variational approach to solve it. It can be rewritten as the sum of
two single particle hamiltonians ($h^{sp}_i$), which describe the
dynamics of the light quarks in the mean field created by the heavy
quark, plus the light--light interaction term, which includes the
Hughes-Eckart  term ($\vec \nabla_1 \cdot \vec\nabla_2 $).
\begin{eqnarray}
H^{\rm int} &=& \sum_{i=q,q^\prime} h^{sp}_i +
V_{qq^\prime}(\vec{r}_1-\vec{r}_2,spin) -
\frac{\vec\nabla_1\cdot\vec\nabla_2}{m_Q} + \sum_{i=q,q^\prime,Q} m_i \\
h^{sp}_i &=& -\frac{\vec\nabla_i^2}{2\mu_i} + V_{Qi}(\vec
r_i,spin), \quad i=q,q^\prime  \label{eq:defhsp}
\end{eqnarray}

\subsection{Quark--Quark Interactions}
\label{sec:int} 
\begin{itemize}
\item {\it Phenomenological Interactions:} We compile here some
phenomenological quark--antiquark potentials fitted to a large sample
of meson states in every flavor sector. The general structure is as
follows ( $i,j=l,s,c,b$):
\begin{equation}
V_{ij}^{q\bar q}(r) = -\frac{\kappa\left ( 1 - e^{-r/r_c}\right )}{r}
+\lambda r^p - \Lambda + \left\{a_0\frac{\kappa}{m_im_j}
\frac{e^{-r/r_0}}{rr_0^2} + \frac{2\pi}{3m_im_j}\kappa^\prime \left (
1 - e^{-r/r_c}\right ) \frac{e^{-r^2/x_0^2}}{\pi^\frac32 x_0^3} \right
\}\vec{\sigma}_i\vec{\sigma}_j  \label{eq:phe}
\end{equation}
with $\vec{\sigma}$ the spin Pauli matrices, $m_i$ the quark masses and
\begin{equation}
x_0(m_i,m_j) = A \left ( \frac{2m_im_j}{m_i+m_j} \right )^{-B} 
\label{eq:phebis}
\end{equation}
We have examined five different interactions, one suggested by Bhaduri
and collaborators~\cite{BD81} (BD) and four suggested by
B. Silvestre-Brac and C. Semay~\cite{Si96,SS93} (AL1, AL2, AP1 y
AP2). 
The parameters of the different potentials are compiled in
Table~\ref{tab:param}. The  potentials considered differ in the form factors
used for the hyperfine terms, the power of the confining term ($p=1$,  as suggested
 by lattice QCD calculations~\cite{GM84}, or $p=2/3$ which for mesons gives the
 correct
 asymptotic Regge trajectories~\cite{Fabre88}), or the use
of a form factor in the OGE Coulomb potential.
%: i) both Yukawa and
%Gaussian, with a mass dependent range, form factors for the hyperfine
%term, ii) two different powers, $p=1$ (suggested by lattice QCD
%calculations~\cite{GM84}) and $2/3$, of the confining term, and iii)
%the effect of a form factor in the OGE Coulomb potential.

The usual $V_{ij}^{qq} = V_{ij}^{q
\bar q}/2$ prescription, coming from a
$\vec{\lambda}_i\vec{\lambda}_j$ color dependence ($\vec{\lambda}$ are
the Gell-Mann matrices) of the whole potential, has been assumed to obtain 
the quark--quark interactions from those given in Eq.~(\ref{eq:phe}).

All those interactions were used in Ref~\cite{Si96} to obtain, in a Faddeev
calculation, the spectrum and static properties of heavy baryons. Our purpose
is to show that our simpler variational method gives equally good results for
all observables analyzed in \cite{Si96} and, besides, it provides us with manageable
wave functions that can be used for the evaluation of further observables.
\begin{table}
\begin{center}
\begin{tabular}{c|lllll}
                          & BD     & AL1    & AL2    & AP1     & AP2  \\\hline
$\kappa$                  & 0.52   & 0.5069 & 0.5871 & 0.4242  & 0.5743  \\
$r_c$[GeV$^{-1}$]         & 0.     & 0.     & 0.1844 & 0.      & 0.3466 \\ 
$p$                       & 1      & 1      &   1    & 2/3     & 2/3   \\
$\lambda$ [GeV$^{(1+p)}$] & 0.186  & 0.1653 & 0.1673 & 0.3898  & 0.3978 \\ 
$\Lambda$ [GeV]           & 0.9135 & 0.8321 & 0.8182 & 1.1313  &
1.1146 \\
$a_0$                     & 1      &   0    &   0    &   0     &   0 \\
$r_0$ [GeV$^{-1}$]        & 2.305  &   $-$  &   $-$  &  $-$    & $-$ \\  
$\kappa^\prime$           & 0.     & 1.8609 & 1.8475 & 1.8025  &
1.8993 \\
$m_u=m_d$ [GeV]           & 0.337  & 0.315  & 0.320  & 0.277   & 0.280
\\
$m_s$  [GeV]              & 0.600  & 0.577  & 0.587  & 0.553   & 0.569
\\
$m_c$  [GeV]              & 1.870  & 1.836  & 1.851  & 1.819   &
1.840\\
$m_b$ [GeV]               & 5.259  & 5.227  & 5.231  & 5.206   & 5.213\\
$B$                       & $-$    & 0.2204 & 0.2132 & 0.3263  &
0.3478 \\
$A$ [GeV$^{B-1}$]         & $-$    & 1.6553 & 1.6560 & 1.5296  &
1.5321\\\hline  
\end{tabular}
\end{center}
\caption{Parameters, from Refs.~\protect\cite{BD81,Si96}, of different
phenomenological potentials (Eqs.~(\protect\ref{eq:phe})
and~(\protect\ref{eq:phebis})).}
\label{tab:param}
\end{table}

\item {\it Chiral Quark Cluster Interactions:} Working in the context
of the SU(2) linear sigma model, Fern\'andez and collaborators have
derived in Ref.~\cite{Fe93} a quark--quark interaction that, apart from the
effective OGE and confining terms, contains a
pseudoscalar ($V^{\rm PS}$) and a scalar ($V^{\rm S}$) potentials
provided by the exchange of Goldstone bosons.  This model was used in
Ref.~\cite{BFV99} to analyze the strangeness-less meson spectrum with good 
results. We analyze this potential in order to check the predictions of the 
SU(2) chirally inspired
quark--quark interactions in the strangeness-less heavy baryon sector.
The
study of  baryons with strangeness would require to extend the model of
Ref.~\protect\cite{Fe93} to three flavors. A linear realization of
SCSB will involve, not only the exchange of new Goldstone bosons, as
kaon or eta mesons, but also more than one scalar meson.
We are not aware of any published potential with such characteristics and 
that performs well in the meson sector. 
Therefore, we prefer to work in this latter case with baryons without strange
quark content.

In the quark--quark sector for
$u$ and $d$ flavors, the potential reads:
\begin{eqnarray}
V_{ij}^{q q} &=& V^{\rm OGE}_{ij} + V^{\rm CON}_{ij} + V^{\rm PS}_{ij}
+ V^{\rm S}_{ij}, \qquad  i,j=u,d \label{eq:cs}\\ 
V^{\rm OGE}_{ij}(r)&=& \frac{\alpha_s}{4}
\vec{\lambda}_i\vec{\lambda}_j \left \{
\frac{1}{r}-\frac{\pi}{m_im_j}\left(1+\frac23
\vec{\sigma}_i\vec{\sigma}_j \right) \frac{e^{-r/r0}}{4\pi r_0^2 r}
\right \} \\ 
V^{\rm CON}_{ij}(r) &=&  \left (-a_c r + a_b \right ) 
\vec{\lambda}_i\vec{\lambda}_j   \\ 
V^{\rm PS}_{ij}(r) &=&
 \frac{\alpha_{ch} m_\pi}{1-m_\pi^2/\Lambda^2_{\rm CSB}} \frac13 \left
\{ Y(m_\pi r) - \frac{\Lambda^3_{\rm CSB}}{m_\pi^3}\, Y(\Lambda_{\rm
CSB}r)\right \} \left (\vec{\sigma}_i\vec{\sigma}_j \right )\left
(\vec{\tau}_i\vec{\tau}_j \right ) \label{eq:ope}\\
V^{\rm S}_{ij}(r) &=&
-\frac{4\alpha_{ch} m_i m_j m_\sigma}{m_\pi^2}
\frac{1}{1-m_\sigma^2/\Lambda^2_{\rm CSB}} \left
\{ Y(m_\sigma r) - \frac{\Lambda_{\rm CSB}}{m_\sigma}\, Y(\Lambda_{\rm
CSB}r)\right \} \label{eq:ult}
\end{eqnarray}
with $\vec{\tau}$ the isospin Pauli matrices, $Y(x) = e^{-x}/x$ and
$\left ( \vec{\lambda}_i\vec{\lambda}_j \right )$ takes the value
$-8/3$ for $qq$ pairs in a baryon. The parameters given in
Refs.~\cite{BFV99,Fe93} are: $r_0=0.145$ fm, $\alpha_s=0.7$,
$\Lambda_{\rm CSB}=3.15$ fm$^{-1}$, $a_c =140$ MeV/fm, $m_u=m_d=313$
MeV, $\alpha_{ch} = 0.027569$, $m_\pi=138$ MeV and $m_\sigma=675$ MeV.
To reproduce the experimental
$\rho-\pi$ mass splitting ($\approx 771-138=633$ MeV), the parameter $r_0$
needs to be slightly reduced to 0.1419 fm.  Besides, the overall
origin of energies, $a_b$, takes the value of 122.1 MeV. Note that
$\left ( \vec{\lambda}_i\vec{\lambda}_j \right )$ takes the value
$-16/3$ for $q\bar q$ pairs in a meson and that the isospin structure
$\left ( \vec{\tau}_i\vec{\tau}_j \right )$ should be multiplied by a
factor $-1$ in the $q\bar q$ case. 

To use the above interaction to predict strangeness-less
heavy baryon masses, such
potential needs to be supplemented by interactions between heavy
quarks ($c$ and $b$) and light quarks ($u$ and $d$). If we look at the
phenomenological potentials discussed in Eq.~(\ref{eq:phe}) and
Table~\ref{tab:param}, we see the AL1 interaction uses a light quark
mass very close to that used in Refs.~\cite{BFV99,Fe93} (315 
versus 313 MeV) and furthermore in both cases a linear confining
potential is used. Besides, the AL1 potential provides masses for the
$D,D^*,B,B^*$ quite close to the experimental values, and  $\pi$ and
$\rho$ masses (138 and 770 MeV respectively) in good agreement with
the values obtained from the interaction of Eq.~(\ref{eq:cs}). Thus,
we will compute the masses of the $\Lambda_{c,b}, \Sigma_{c,b}$ and
$\Sigma_{c,b}^*$ baryons with a new potential, which we  call
AL1$\chi$, defined as follows:
\begin{itemize}
\item The light--light quark ($u$ and $d$) interactions will be
described by the potential of Eq.~(\ref{eq:cs}), which incorporates a
pattern of SCSB, and the parameters given below 
Eq.~(\ref{eq:ult}), but with $m_u=m_d=315$ MeV, $r_0=0.1414$ fm and
$a_b=122.4$ MeV. These two latter parameters have been very slightly
modified to obtain 138 and 770 MeV (values provided by the AL1
potential) for the $\pi$ and $\rho$ masses respectively, when a $u,d$
quark mass of 315 MeV is used.  The new set of parameters provides
states with masses of 138, 1236 and 1918 (770, 1558 and 2173) MeV with
$\pi$ ($\rho$) quantum numbers, in perfect agreement with the results
of Ref.~\cite{BFV99}.

\item To describe the interaction between the heavy quarks and the $u$ and $d$
quarks we will adopt the AL1 model.
 
\end{itemize}
In this way, and by comparing results obtained from both the AL1 and
AL1$\chi$ interactions, we will be able to test the effects of the
inclusion of a SCSB pattern to describe the light--light quark
interaction in the $u,d$ sector. Such a study, though it has already
been performed in other systems, as light meson or baryon
spectrum, etc, has not been ever carried out for heavy
baryons. It is of great interest since the quantum numbers of the
light degrees of freedom in a $\Lambda-$ or $\Sigma-$type heavy
baryon have a clear correspondence to those of the $\pi$ or $\rho$ meson.

\end{itemize}

\section{Variational Wave Functions}\label{vwf}
In a baryon, the singlet color wave function is completely
anti-symmetric under the exchange of any of the three quarks. Within
the SU(3) quark model, we assume a complete symmetry of the wave
function under the exchange of the two  light quarks ($u,d,s$) flavor,
spin and space degrees of freedom. On
the other hand, for the interactions described in the previous
subsection, we have that both the total spin of the baryon, ${\vec
S}_{\rm B} = \left ( {\vec \sigma}_q + {\vec \sigma}_{q^\prime} +
{\vec \sigma}_Q \right )/2$, and the orbital angular momentum of the
light quarks with respect to $Q$, $\vec{L}$, defined as
\begin{equation}
{\vec L} = {\vec l}_1 + {\vec l}_2, \qquad {\rm with}~~ {\vec
l}_k=-{\rm i}~~{\vec r}_k \times {\vec \nabla}_k, \quad k=1,2
\end{equation}
commute with the intrinsic hamiltonian. We will assume that the
ground states of the baryons are in s--wave, $L=0$. This implies that
the spatial wave function can only depend on the
relative distances $r_1$, $r_2$ and
$r_{12}=|\vec{r}_1-\vec{r}_2|$. Note that when the heavy quark mass is
infinity ($m_Q \to \infty$), the total spin of the light degrees of
freedom, $\vec{S}_{\rm light} = \left ( {\vec \sigma}_q + {\vec
\sigma}_{q^\prime} \right )/2$, commutes with the intrinsic
hamiltonian, since the
$\vec{\sigma}_Q\cdot\vec{\sigma}_{q,q^\prime}/(m_Qm_{q,q^\prime})$ terms
vanish in this limit. With all these ingredients, and taking into account the
quantum numbers of the light degrees of freedom for each baryon,
compiled in Table~\ref{tab:summ} and that in general are always well
defined in the static limit mentioned above, we have used the
following wave functions in our variational approach\footnote{An
obvious notation has been used for the isospin--flavor ($|I,M_I\rangle _I$,
$|ls\rangle $ or $|sl\rangle $) and spin ($|S,M_S\rangle _{S_{\rm light}}$) wave functions
of the light degrees of freedom.}
\begin{itemize}
\item {\it $\Lambda-$type baryons: $I=0,~S_{\rm light}=0$}
\begin{eqnarray}
|\Lambda_Q; J=\frac12, M_J\left. \right \rangle  &=& \Big \{  |00 \rangle _I 
\otimes  |0 0 \rangle _{S_{\rm light}} \Big \} 
\Psi_{ll}^{\Lambda_Q} (r_1,r_2,r_{12}) \otimes  |Q; M_J\rangle  \label{eq:sim}
\end{eqnarray}
where $\Psi_{ll}^{\Lambda_Q} (r_1,r_2,r_{12}) = \Psi_{ll}^{\Lambda_Q}
(r_2,r_1,r_{12})$ to guarantee a complete symmetry of the wave function
under the exchange of the two light quarks ($u,d$) flavor, spin and
space degrees of freedom, and finally $M_J$ is the baryon total
angular momentum third component. Note, that SU(3) flavor symmetry (SU(2), in
the case of the $\Lambda_Q$ baryon) would also 
allow for a component in the wave function of the type
\begin{equation}
 \sum_{M_SM_Q} (\frac12 1 \frac12 | M_QM_SM_J)\Big \{  |00 \rangle _I 
\otimes  |1 M_S \rangle _{S_{\rm light}} \Big \} 
 \Theta_{ll}^{\Lambda_Q} (r_1,r_2,r_{12}) \otimes  |Q; M_Q\rangle  \label{eq:antis}
\end{equation}
with $\Theta_{ll}^{\Lambda_Q} (r_1,r_2,r_{12}) =
-\Theta_{ll}^{\Lambda_Q} (r_2,r_1,r_{12})$ (for instance terms of the
type $r_1-r_2$) and, the real numbers $(j_1j_2j|m_1m_2m)=\langle
j_1m_1j_2m_2|jm\rangle $ are Clebsh-Gordan coefficients. This
component is forbidden by HQS in the limit $m_Q \to \infty$, where
$S_{\rm light}$ turns out to be well defined and set to zero for
$\Lambda_Q-$type baryons. The most general SU(2) $\Lambda_Q$ wave
function will involve a linear combination of the two components,
given in Eqs.~(\ref{eq:sim}) and (\ref{eq:antis}). Neglecting ${\cal
O}(q/m_Q)$, HQS imposes an additional constraint, which justifies the
use of a wave function of the type of that given in Eq.~(\ref{eq:sim})
with the obvious simplification of the three body problem.

\item {\it $\Sigma$ and $\Sigma^*-$type baryons: $I=1,~S_{\rm light}=1$}. 

\begin{eqnarray}
|\Sigma_Q; J=\frac12, M_J; M_T\left. \right \rangle  &=&
\sum_{M_SM_Q}(\frac12 1 \frac12| M_Q M_S M_J) \Big \{ |1 M_T\rangle _I
|\otimes |1 M_S\rangle _{S_{\rm light}} \Big \} 
\nonumber  \\
&\times & \Psi_{ll}^{\Sigma_Q}(r_1,r_2,r_{12})\otimes|Q; M_Q\rangle  \label{eq:sig}\\ 
|\Sigma_Q^*;
|J=\frac32, M_J; M_T\left. \right \rangle  &=& \sum_{M_SM_Q} (\frac12 1
\frac32 |M_Q M_S M_J) \Big \{ |1 M_T\rangle _I \otimes | 1 M_S\rangle _{S_{\rm
|light}} \Big \} \nonumber \\
&\times & \Psi_{ll}^{\Sigma_Q^*}(r_1,r_2,r_{12}) \otimes |Q;M_Q\rangle 
\end{eqnarray}
with $M_T$ and $M_Q$, the baryon isospin and heavy quark spin third
components, respectively. 

As in the $\Lambda_Q$ case, and based on the HQS predictions, here we
have also neglected components constructed out of the spin--singlet ($|0
0 \rangle _{S_{\rm light}}$) and anti-symmetric spatial wave functions, since
they are suppressed by powers of $1/m_Q$. HQS
leads to similar simplifications for the rest of baryons studied below
and we will omit any further comment in what follows.

\item {\it $\Xi-$type baryons: $I=\frac12, S_{\rm light}=0$}
\begin{eqnarray}
|\Xi_Q; J=\frac12, M_J; M_T\left. \right \rangle  &=& \frac{1}{\sqrt2}\Big\{ 
|ls\rangle \Psi_{ls}^{\Xi_Q}(r_1,r_2,r_{12}) -
|sl\rangle \Psi_{sl}^{\Xi_Q}(r_1,r_2,r_{12}) \Big \} \otimes |00\rangle _{S_{\rm
light}} 
\otimes |Q; M_J\rangle 
\end{eqnarray}
where the isospin third component of the baryon, $M_T$, is that of the light
quark $l$ ($1/2$ or $-1/2$ for the $u$ or the $d$ quark, respectively). 

\item {\it $\Xi^\prime$ and $\Xi^*-$type baryons: $I=\frac12,~S_{\rm
light}=1$}.

\begin{eqnarray}
|\Xi^\prime_Q; J=\frac12, M_J; M_T\left. \right \rangle  &=& \sum_{M_SM_Q} (\frac12 1
\frac12| M_Q M_S M_J) \frac{1}{\sqrt2}  \nonumber\\
&\times&\Big \{ |ls\rangle \Psi_{ls}^{\Xi^\prime_Q}(r_1,r_2,r_{12}) +
|sl\rangle \Psi_{sl}^{\Xi^\prime_Q}(r_1,r_2,r_{12}) \Big \} \otimes |1 M_S\rangle _{S_{\rm light}}\otimes |Q; M_Q\rangle  \\
|\Xi^*_Q; J=\frac32, M_J; M_T\left. \right \rangle  &=& \sum_{M_SM_Q}(\frac12 1
\frac32| M_Q M_S M_J) \frac{1}{\sqrt2}  \nonumber\\
&\times&\Big \{ |ls\rangle \Psi_{ls}^{\Xi^*_Q}(r_1,r_2,r_{12}) +
|sl\rangle \Psi_{sl}^{\Xi_Q^*}(r_1,r_2,r_{12}) \Big\} \otimes |1 M_S\rangle _{S_{\rm light}}
 \otimes |Q; M_Q\rangle 
\end{eqnarray}
where the isospin third component of the baryon,$M_T$, is that of the light
quark $l$.

\item {\it $\Omega$ and $\Omega^*-$type baryons: $I=0,~S_{\rm light}=1$}. 
\begin{eqnarray}
|\Omega_Q; J=\frac12, M_J \left. \right \rangle  &=& \sum_{M_SM_Q} (\frac12 1 \frac12
|M_Q M_S M_J) |1 M_S\rangle _{S_{\rm light}} 
 \Psi_{ss}^{\Omega_Q}(r_1,r_2,r_{12}) \otimes |Q; M_Q\rangle  \\ |\Omega_Q^*;
J=\frac32, M_J \left. \right \rangle  &=& \sum_{M_SM_Q} (\frac12 1 \frac32 |M_Q M_S
M_J) | 1 M_S\rangle _{S_{\rm light}}\Psi_{ss}^{\Omega^*_Q}(r_1,r_2,r_{12})
|\otimes |Q; M_Q\rangle  \label{eq:ome*}
\end{eqnarray}
\end{itemize}
where  $\Psi_{ss}^{\Lambda_Q} (r_1,r_2,r_{12}) = \Psi_{ss}^{\Lambda_Q}
(r_2,r_1,r_{12})$.

The spatial wave function, $\Psi_{qq^\prime}^{B_Q}$, will be
determined by the variational principle: $\delta \langle B_Q | H^{\rm
int}| B_Q \rangle = 0$. For simplicity, we will use a family of
functions with free parameters, which will be determined by minimizing
the mass of the baryon. The OGE has a regularized delta function,
which provides strong short range repulsions/attractions, specially
between the light quarks\footnote{The size of this term of the OGE
interaction is proportional to the inverse of the masses of the
involved quarks.}. A similar situation appears in the context of
double $\Lambda$ hypernuclei~\cite{Ca99}, which in a very good
approximation can be also reduced to a three body problem.  In this
context, it was shown that the use of a Jastrow--type functional form
for the spatial wave functions leads to excellent results, and to
notably simpler wave functions than other functional forms, as for
instance a series of standard Hylleraas type wave
functions~\cite{Al02}.  Thus, we take
\begin{eqnarray}
\Psi_{qq^\prime}^{B_Q} (r_1,r_2,r_{12}) &=& N F^{B_Q}(r_{12})
\phi_q^Q(r_1)\phi_{q^\prime}^Q(r_2)
\end{eqnarray}
where $N$ is a constant, which is determined from
normalization\footnote{Note that, the Jastrow form assumes a
factorization of the wave function in three terms, each of them
depends only on one of the involved variables $r_1$, $r_2$ and
$r_{12}$. Dependences of the type $r_1-r_2$, mentioned in the
discussion of Eq.~(\ref{eq:antis}), cannot be accommodated with
factorizable functions of this type.}
\begin{eqnarray}
1=\int d^3r_1 \int d^3 r_2 \left |\Psi_{qq^\prime}^{B_Q}(r_1,r_2,r_{12})
\right |^2 = 8\pi^2 \int_0^{+\infty}dr_1~r_1^2~ \int_0^{+\infty}dr_2~r_2^2
\int_{-1}^{+1} d\mu~\left |\Psi_{qq^\prime}^{B_Q}(r_1,r_2,r_{12})
\right |^2 
\end{eqnarray}
where $\mu$ is the cosine of the angle between the vectors $\vec
{r}_1$ and $\vec {r}_2$, being $r_{12}=(~r_1^2+r_2^2-2 r_1 r_2
\mu)^\frac12$. For simplicity, we do not entirely determine the
functions $\phi_q^Q$ and $\phi_{q^\prime}^Q$ from the variational
principle, but we rather fix the bulk of these functions to the
$s-$wave ground states ($\varphi_{i=q,q^\prime}^Q$) of the single
particle hamiltonians, $h^{sp}_{i=q,q^\prime}$, defined in
Eq.~(\ref{eq:defhsp}), and modify their behavior at large
distances. Thus, we take
\begin{eqnarray}
\phi_q^Q (r_1) &=& (1+\alpha_qr_1)\varphi_q^Q(r_1) \nonumber\\
\phi_{q^\prime}^Q (r_2) &=& (1+\alpha_{q^\prime}r_2)\varphi_{q^\prime}^Q(r_2) 
\label{eq:onebody}
\end{eqnarray}
with only one (two) free parameter for a $ll$ or $ss$ ($ls$) baryon
light quark content. Besides, we construct the light--light correlation
function, $F^{B_Q}$, from a linear combination of
gaussians,
\begin{eqnarray}
F^{B_Q}(r_{12}) &=& f^{B_Q}(r_{12})
\sum_{j=1}^4 a_j e^{-b_j^2(r_{12}+d_j)^2},\quad a_1=1 \label{eq:fbig12}\\ 
&&\nonumber \\
f^{B_Q}(r_{12}) &=& \left\{ \begin{array}{lcl}  1 -
e^{-cr_{12}} &{\rm if} & V^B_{qq^\prime} (r_{12}=0) \gg 0 \\
&&\\ 1~~ (c\to + \infty) &{\rm if} & V^B_{qq^\prime}
(r_{12}=0) \le 0
\end{array}\right.\label{eq:f12}
\end{eqnarray}
where $V^B_{qq^\prime}$ denotes the light--light interaction projected
onto the spin and isospin quantum numbers of the baryon $B$. The
correlation function, $F^{B_Q}$, should vanish at large distances
because of the confinement potential\footnote{The confinement
potential is also responsible for the non--vanishing values of the
parameters $\alpha_q$ and $\alpha_q^\prime$ in
Eq.~(\protect\ref{eq:onebody}). Indeed, if at large distances the
light--light interaction vanishes ($V^B_{qq^\prime}(r_{12}\gg 1) \to 0
$) then the product of single particle wave functions
$\varphi_q^Q(r_1)~\varphi_{q^\prime}^Q(r_2)$ will provide the correct
spatial dependence, at large distances, of the solution of the three
body problem.}. The function $f^{B_Q}$ vanishes at the origin in those
cases where $V^B_{qq^\prime}$ is highly repulsive a short distances
(this is the case for instance for the AL1$\chi$ interaction and
$S_{\rm light}=1$, i.e. $\Sigma,\Sigma^*-$type baryons). Note that, one
of the $a_j$ parameters can be absorbed into the normalization
constant $N$, thus besides the $c$ and $\alpha_{q,q^\prime}$
parameters, there are a total of eleven (we fix $a_1=1$) free
parameters to be determined by the variational principle and the mass
of the baryon is just the expected value of the intrinsic
hamiltonian. Note that all interactions of Subsect.~\ref{sec:int} are
diagonal in the flavor space, except for the pion exchange term of the
AL1$\chi$ one, and that the following relationships are of interest
to compute the kinetic energy expected values
\begin{eqnarray}
\vec{\nabla}_1 \Big[\phi_1(r_1)\phi_2(r_2)F(r_{12})\Big] &=& 
\frac{\vec{r}_1}{r_1}\frac{d\phi_1(r_1)}{dr_1}\phi_2(r_2)F(r_{12}) + 
\frac{\vec{r}_1-\vec{r}_2}{r_{12}}\phi_1(r_1)\phi_2(r_2)
\frac{dF(r_{12})}{dr_{12}}\nonumber\\
\vec{\nabla}_2 \Big[\phi_1(r_1)\phi_2(r_2)F(r_{12})\Big] &=& 
\frac{\vec{r}_2}{r_2}\phi_1(r_1)\frac{d\phi_2(r_2)}{dr_2}F(r_{12}) - 
\frac{\vec{r}_1-\vec{r}_2}{r_{12}}\phi_1(r_1)\phi_2(r_2)
\frac{dF(r_{12})}{dr_{12}}\label{eq:nabla}
\end{eqnarray}
The expected value of the Hughes-Eckart contribution ($\vec \nabla_1
\cdot \vec\nabla_2 $) to the kinetic energy is readily obtained from
the above equations after integrating by parts.

Given the spatial part of the baryon wave function,
$\Psi_{qq^\prime}^{B_Q}(r_1,r_2,r_{12})$, the probability, ${\cal
P}_l$, of finding each of the two light quarks with angular momentum
$l$, referred to the heavy quark, and coupled to $L=0$ is given by
\begin{eqnarray}
{\cal P}_l & = & 4 \pi^2 (2 l +1
)~\int_0^{+\infty}dr_1~r_1^2~ \int_0^{+\infty}dr_2~r_2^2~\left |
\int_{-1}^{+1}d\mu~P_{l}(\mu) \Psi_{qq^\prime}^{B_Q}(r_1,r_2,r_{12})
\right |^2, \label{eq:mult}
\end{eqnarray}
where $P_l$ is the Legendre Polynomial of order $l$. 

\section{Static Properties: Mass and Charge Densities And
Orbital Magnetic Moments}\label{sec:sp}

Since the charge operator is diagonal in the spin-flavor space, the
baryon charge density at the point $P$ (coordinate vector $\vec{r}$ in 
the CM frame, see Fig.~\ref{fig:coor}) is given by: 
\begin{eqnarray}
\rho_e^{B_Q} (\vec{r}) &=& \int d^3R~ d^3 r_1 d^3 r_2 \Big |
\frac{e^{{\rm i}\vec{P}_{CM}\vec{R}}}{\sqrt
V}\Psi_{qq^\prime}^{B_Q}(r_1,r_2,r_{12}) \Big |^2 \left \{ e_Q
\delta^3 (\vec{r}-\vec{y}_h) + e_q \delta^3 (\vec{r}-\vec{y}_1) +
e_{q^\prime} \delta^3 (\vec{r}-\vec{y}_2) \right \}
\nonumber \\ 
& = & \int d^3 r_1 d^3 r_2 \Big |
\Psi_{qq^\prime}^{B_Q}(r_1,r_2,r_{12}) \Big |^2 \left \{ e_Q \delta^3
(\vec{r}-\vec{y}_h) + e_q \delta^3 (\vec{r}-\vec{y}_1) + e_{q^\prime}
\delta^3 (\vec{r}-\vec{y}_2) \right \} \nonumber \\ 
& \equiv & \rho_e^{B_Q} (\vec{r})\big|_Q 
+ \rho_e^{B_Q}(\vec{r})\big|_q  +
\rho_e^{B_Q}(\vec{r})\big|_{q^\prime} \label{eq:dens}
\end{eqnarray} 
where $\delta (\cdots)$ and $\vec{P}_{CM}$ are the three-dimensional
Dirac's delta and the total baryon momentum respectively, $V$ is the
interacting volume\footnote{It cancels out with $\int d^3R$.},
$e_{q,q^\prime,Q}$ are the quark charges, in proton charge units
($e$), and from Fig.~\ref{fig:coor} we have\footnote{There exists the
obvious restriction
$m_Q\vec{y}_h+m_q\vec{y}_1+m_{q^\prime}\vec{y}_2=0$.}, $\vec{y}_h= -
\left (m_q\vec{r}_1+m_{q^\prime}\vec{r}_2 \right)/M$,
$\vec{y}_1=\vec{y}_h+\vec{r}_1$ and
$\vec{y}_2=\vec{y}_h+\vec{r}_2$. The charge density is spherically
symmetric ($\rho_e^{B_Q} (\vec{r}) = \rho_e^{B_Q} (|\vec{r}~|)$),
since the wave function only depend on scalars ($r_1,r_2$ and
$\vec{r}_1\cdot \vec{r}_2$) for $L=0$ baryons. The charge form factor
is defined as usual
\begin{eqnarray}
{\cal F}_e^{B_Q} (\vec{q}~) = \int d^3 r e^{{\rm
i}\vec{q}\cdot\vec{r}}\rho_e^{B_Q} (r)\label{eq:fe}
\end{eqnarray}
and it only depends  on $|\vec{q}~|$. The
charge mean square radii are defined
\begin{equation}
\langle r^2 \rangle_e^{B_Q} = 
\int d^3 r r^2 \rho_e^{B_Q} (r) = 
4\pi \int_0^{+\infty}dr r^4 \rho_e^{B_Q} (r) 
\label{eq:r2q}
\end{equation}
The baryon mass density, $\rho_m^{B_Q} (r)$ is
readily obtained from Eq.~(\ref{eq:dens}) with the obvious
substitutions $(e_q,e_{q^\prime},e_Q) \to (m_q/M,m_{q^\prime}/M,m_Q/M)$. 
\begin{eqnarray}
\rho_m^{B_Q} (\vec{r}) &=&
\int d^3 r_1 d^3 r_2 \Big |
\Psi_{qq^\prime}^{B_Q}(r_1,r_2,r_{12}) \Big |^2 \left \{ \frac{m_Q}{M} \delta^3
(\vec{r}-\vec{y}_h) + \frac{m_q}{M} \delta^3 (\vec{r}-\vec{y}_1) + 
\frac{m_{q^\prime}}{M}
\delta^3 (\vec{r}-\vec{y}_2) \right \} \nonumber \\ 
& \equiv & \rho_m^{B_Q} (\vec{r})\big|_Q 
+ \rho_m^{B_Q}(\vec{r})\big|_q  +
\rho_m^{B_Q}(\vec{r})\big|_{q^\prime} \label{eq:mdens}
\end{eqnarray}
where we have normalized $\rho_m^{B_Q} (r)$ to 1. The
mass mean square radii are defined
\begin{equation}
\langle r^2 \rangle_m^{B_Q} = \int d^3 r r^2 \rho_m^{B_Q} (r) = 
4\pi \int_0^{+\infty}dr r^4 \rho_m^{B_Q} (r) 
\label{eq:r2m}
\end{equation}
The intrinsic orbital magnetic moment is defined in terms of the
velocities $\vec{v}_{h,1,2}$ of the quarks $Q,q$ and $q^\prime$, 
with respect to the position of the CM, and it reads
\begin{eqnarray}
\vec{\mu}^{B_Q} &=& \int d^3R~ d^3 r_1 d^3 r_2 \frac{e^{-{\rm
i}\vec{P}_{CM}\cdot\vec{R}}}{\sqrt
V}\left(\Psi_{qq^\prime}^{B_Q}(r_1,r_2,r_{12})\right)^* \left \{
\frac{e_Q}{2m_Q} (\vec{y}_h \times m_Q\vec{v}_{y_h}) \right.\nonumber \\
&+&\left. \frac{e_q}{2m_q} (\vec{y}_1 \times m_q\vec{v}_{y_1}) + 
\frac{e_{q^\prime}}{2m_{q^\prime}} (\vec{y}_2 \times m_{q^\prime}
\vec{v}_{y_2}) \right \}\frac{e^{{\rm
i}\vec{P}_{CM}\cdot\vec{R}}}{\sqrt
V}\Psi_{qq^\prime}^{B_Q}(r_1,r_2,r_{12})
\end{eqnarray}
with\footnote{Note that the classical kinetic energy  has a term on 
$\vec{v}_{y_1}\cdot\vec{v}_{y_2}$ and then the operator
$m_q \vec{v}_{y_1}$ is not proportional to 
$-i\stackrel{\to}{\nabla}_{y_1}$, but it is rather given by
$m_q \vec{v}_{y_1}=\frac{M-m_q}{ M}(-i\stackrel{\to}{\nabla}_{y_1}) 
-\frac{m_{q^\prime}}{M}(-i\stackrel{\to}{\nabla}_{y_2})
=(-i\stackrel{\to}{\nabla}_1)$. Similarly
 $m_{q^\prime} \vec{v}_{y_2}=(-i\stackrel{\to}{\nabla}_2)$.}
 $m_q\vec{v}_{y_1} = - i \vec{\nabla}_1$,
$m_{q^\prime}\vec{v}_{y_2} = - i \vec{\nabla}_2$ and $m_Q\vec{v}_{y_h}
= i \left ( \vec{\nabla}_1 + \vec{\nabla}_2 \right)$. Since the wave
function only depends on scalars for a $L=0$ baryon, the intrinsic
orbital magnetic moment vanishes. Note that the intrinsic orbital magnetic
moment is zero despite the quark pairs $(Q,q)$ and $(Q,q^\prime)$ are
not in relative $s-$waves, as mentioned above in the discussion of
Eq.~(\protect\ref{eq:mult}). Thus, the magnetic moment of the baryon
is entirely due to the spin contribution and since we are adopting HQS
wave functions where the light degrees of freedom spin wave function
factorizes, the magnetic moment should coincide to that
obtained in the naive quark-model treatment of the baryon.

Finally and as a further test of our variational wave--functions, we
compute some quantities which in Ref.~\cite{Si96} are denoted by
$\rho_i(0)$ and called wave function at the origin for the ($jk$)
pair, where $(i,j,k)$ is a cyclic permutation of ($Q,q,q^\prime$). In this 
reference the following Jacobi coordinates 
\begin{eqnarray}
\vec{s}_Q (\vec{r}_1,\vec{r}_2) &=& \vec{r}_1-\vec{r}_2\nonumber\\
\vec{s}_q (\vec{r}_1,\vec{r}_2) &=& a_q \vec{r_2}, \quad a_q = 
\sqrt{\frac{m_Q(m_q+m_{q^\prime})}{m_q(m_Q+m_{q^\prime})}}\nonumber\\
\vec{s}_{q^\prime} (\vec{r}_1,\vec{r}_2) &=& a_{q^\prime} \vec{r_1}, 
\quad a_{q^\prime} = 
-\sqrt{\frac{m_Q(m_q+m_{q^\prime})}{m_{q^\prime}(m_Q+m_q)}}
\end{eqnarray}
and correlation functions\footnote{These correlation functions satisfy
$\int_0^{+\infty}ds s^2 \eta_i^{B_Q}(s) = 1$.}
\begin{equation}
\eta_i^{B_Q}(s) = \frac{1}{s^2}\int d^3r_1 d^3r_2 \Big |
\Psi_{qq^\prime}^{B_Q}(r_1,r_2,r_{12}) \Big |^2
\delta\left(s-\big |\vec{s}_i(\vec{r}_1,\vec{r}_2)\big |\right), 
\quad i=Q,q,q^\prime
\end{equation}
are introduced. Thus, the wave functions at the origin are defined
by~\cite{Si96}
\begin{eqnarray}
\eta_Q^{B_Q}(s=0) &=& 16 \pi^2 \int_0^{+\infty} dt t^2
\Big | \Psi_{qq^\prime}^{B_Q}\left(r_1=t,r_2=t,r_{12}=0\right) \Big
|^2 \nonumber \\ 
\eta_q^{B_Q}(s=0) &=& \frac{16
\pi^2}{a_q^3} \int_0^{+\infty} dt t^2 \Big |
\Psi_{qq^\prime}^{B_Q}\left(r_1=t,r_2=0,r_{12}=t\right) \Big |^2
\nonumber \\ 
\eta_{q^\prime}^{B_Q}(s=0) &=& \frac{16
\pi^2}{|a_{q^\prime}^3|} \int_0^{+\infty} dt t^2 \Big |
\Psi_{qq^\prime}^{B_Q}\left(r_1=0,r_2=t,r_{12}=t\right) \Big |^2
\label{eq:eta0}
\end{eqnarray}

\section{Results}\label{sec:res}
In Table~\ref{tab:masses} we present variational results for the
 charmed and bottom baryon masses obtained with each of the six
 quark--quark interactions considered in this work. We use an
 integration step of $10^{-2}$ fm and we estimate in $0.5$ MeV the
 numerical error of the variational masses given in the table.  On the
 other hand, by including a higher number of gaussians\footnote{In the
 Appendix (Table~\ref{tab:ng}), we illustrate, for the $\Lambda_c$
 baryon with an AL1 inter-quark interaction, the dependence of the VAR
 results on the number of gaussians.}  in the light--light correlation
 function (Eq.~(\ref{eq:fbig12})), or quadratic or cubic terms in the
 one body wave--functions (Eq.~(\ref{eq:onebody})), the variational
 energies cannot be lowered in more than $0.5$ MeV.  The corresponding
 wave functions can be reconstructed from the parameters given in
 Tables~\ref{tab:charmparam} and \ref{tab:bottomparam} in the
 Appendix. We use a {\it Simplex} algorithm to find out the position
 of the minima~\cite{numrec} and typically for each baryon the
 minimization procedure, including the construction of the wave
 function, takes around twenty minutes in a 2 GHz Pentium IV
 processor.

 Besides, in Table~\ref{tab:masses}, we compare, when
possible, with the Faddeev results of Ref.~\cite{Si96}. We find an
excellent agreement between our variational results and those of
Ref.~\protect\cite{Si96}. In most cases we find discrepancies
smaller\footnote{Such differences are certainly small fractions of the
baryon binding energies.} than 3 MeV, and only in three cases (BD
$\Omega_c$, $\Sigma_b$ and AL2 $\Lambda_b$ baryons) Faddeev and
variational masses differ in more than 5 MeV. Faddeev masses are
usually smaller than variational ones, but in some cases the
variational approach presented here provides lower values, and thus
better estimates of the baryon masses (see f.i. AL2 $\Lambda_b$, where
we find a mass of around 8 MeV smaller than that quoted in
Ref.~\cite{Si96}).  Having in mind that the Faddeev masses suffer from
numerical uncertainties of the order of 5 MeV~\protect\cite{Si96}, we
can safely conclude that our variational approach reproduces the
ground state masses quoted in Ref.~\protect\cite{Si96}.  Thus,
corrections to the leading HQS wave functions (Eq.~(\ref{eq:sim}) and
Eqs.~(\ref{eq:sig})--(\ref{eq:ome*})) should provide quite small (let
us say, smaller than 5 MeV) changes in ground state masses, even for the charm
sector.

\begin{table}
\begin{tabular}{lc|ccccccl|ccccccl}
            &&\multicolumn{7}{c}{Charm}&\multicolumn{7}{c}{Bottom} 
\\\tstrut
Baryon      &     & BD   & AL2  & AP1  & AP2  & AL1  & AL1$\chi$ &~Exp~
                  & BD   & AL2  & AP1  & AP2  & AL1  & AL1$\chi$ &~Exp~
\\\hline\tstrut
            & FAD & 2321 & 2302 & 2311 & 2313 & 2296 &   $-$ &
                  & 5660 & 5637 & 5659 & 5643 & 5643 &   $-$ &      \\
~~~$\Lambda$& VAR & 2320 & 2305 & 2308 & 2310 & 2295 & 2140  & $2285\pm 1 $
                  & 5663 & 5629 & 5659 & 5643 & 5643 & 5475  & $5624\pm 9$\\\hline
            & FAD & 2494 & 2474 & 2484 & 2491 & 2466 &   $-$ &
                  & 5865 & 5844 & 5868 & 5859 & 5849 &   $-$ &      \\
~~~$\Sigma$ & VAR & 2498 & 2479 & 2484 & 2491 & 2469 & 2457  & $2452\pm 1$  
                  & 5871 & 5844 & 5870 & 5861 & 5851 & 5841  & $5770^\dagger\pm 70$    \\\hline
~~~$\Sigma^*$& VAR& 2570 & 2559 & 2560 & 2571 & 2548 & 2535  & $2518\pm 2$ 
                  &  5900& 5874 & 5899 & 5893 & 5882 & 5869  & $5780^\dagger\pm 70$   \\\hline
            & FAD & 2502 & 2480 & 2485 & 2488 & 2473 &   $-$ &
                  & 5830 & 5803 & 5819 & 5805 & 5808 &   $-$ &      \\
~~~$\Xi$    & VAR & 2506 & 2480 & 2485 & 2488 & 2474 &   $-$ & $2469\pm 3$ 
                  & 5832 & 5800 & 5820 & 5807 & 5808 &   $-$ & $5760^\dagger\pm 70$ \\\hline
~~~$\Xi^\prime$&VAR& 2606 & 2584 & 2587 & 2592 & 2578 &  $-$ & $2576\pm 3$
                  & 5963 & 5939 & 5959 & 5946 & 5946 &   $-$ & $5900^\dagger\pm 70$     \\\hline
~~~$\Xi^*$  & VAR & 2671 & 2661 & 2664 & 2672 & 2655 &   $-$ & $2646\pm 2$
                  & 5989 & 5970 & 5987 & 5979 & 5975 &   $-$ & $5900^\dagger\pm 80$     \\\hline
            & FAD & 2707 & 2686 & 2678 & 2682 & 2678 &   $-$ &
                  & 6048 & 6029 & 6033 & 6020 & 6035 &   $-$ &      \\
~~~$\Omega$ & VAR & 2713 & 2683 & 2682 & 2684 & 2681 &   $-$ & $2698\pm 3$
                  & 6050 & 6030 & 6036 & 6023 & 6033 &   $-$ & $5990^\dagger\pm 70$ \\\hline
~~~$\Omega^*$& VAR& 2770 & 2759 & 2759 & 2765 & 2755 &   $-$ & $2660^\dagger\pm 80$ 
                  & 6074 & 6061 & 6068 & 6056 & 6063 &   $-$ & $6000^\dagger\pm 70$     \\\hline
\end{tabular}
\caption{Masses (in MeV) of the ground states of different baryons 
with the six potentials considered in this
work. Variational and Faddeev results are given in the rows denoted by
VAR and FAD, respectively. The latter ones are taken from
Ref.~\protect\cite{Si96}, but the effect of the three body forces
considered in that reference, $V_{123}= {\rm
constant}/m_qm_{q^\prime}m_Q$, has been eliminated.  Experimental
masses are taken from Ref.~\protect\cite{pdg02} when possible, and the
lattice QCD estimates of Ref.~\protect\cite{bowler} are quoted for
those cases ($\dagger$) where the experimental masses are not known.}
\label{tab:masses}
\end{table}

 We have also computed the $\Sigma^*$, $\Xi^\prime$, $\Xi^*$ and $\Omega^*$
ground state masses, which were not evaluated in the work of
Ref.~\protect\cite{Si96}. Note that, the
$\vec{\sigma}_q\cdot\vec{\sigma}_Q/m_qm_Q$ OGE term
breaks\footnote{For s--wave baryons, the spin operator
$\vec{\sigma}_q\cdot\vec{\sigma}_Q$ vanishes for $S_{\rm
light}=0$ wave functions and for spin triplet states, $S_{\rm
light}=1$,  takes the values $1$ and $-2$, for baryon total angular
momentum $J=\frac32$ and $\frac12$, respectively.} the mass degeneration
between the $S_{\rm light}$ triplet states ($\Sigma$-$\Sigma^*$,
$\Xi^\prime$-$\Xi^*$ and $\Omega$-$\Omega^*$). The five
phenomenological interactions studied in Ref.~\cite{Si96} lead to
ground state masses which compare reasonably well with the existing
experimental data and lattice QCD mass estimates for all baryons.

Besides, we see that while the AL1 and AL1$\chi$ interactions lead to
similar masses (differences of the order of 10 MeV) for the $\Sigma_Q$
and $\Sigma^*_Q$ baryons, the chirally inspired potential,
AL1$\chi$, predicts masses for the $\Lambda_Q$ baryons  of about 155 (charm
sector) and 168 (bottom sector) MeV smaller than those obtained from
the phenomenological AL1 interaction. In terms of binding energies 
($B_Q=M[{\Lambda_Q}]-m_Q-m_u-m_d$), we have relative changes as large
as 90\%,
\begin{equation}
\Big |\frac{B_c^{\rm AL1}-B_c^{{\rm AL1}\chi}}{B_c^{\rm AL1}}\Big | =
0.9, \qquad \Big |\frac{B_b^{\rm AL1}-B_b^{{\rm AL1}\chi}}{B_b^{\rm
AL1}}\Big | = 0.7
\end{equation}
Such big changes are due to the different behaviour of the One Pion
Exchange (OPE) potential in the light quark--quark and
quark--antiquark sectors. In the first case we have, from
Eq.~(\ref{eq:ope}), the operator $\left (\vec{\sigma}_i\vec{\sigma}_j
\right )\left (\vec{\tau}_i\vec{\tau}_j \right )$ which provides a
factor $9$ ($1$) for the light quantum numbers of the $\Lambda_Q$
($\Sigma_Q$ or $\Sigma_Q^*$) baryon. The spin--isospin operator turns
out to be $-\left (\vec{\sigma}_i\vec{\sigma}_j \right )\left
(\vec{\tau}_i\vec{\tau}_j \right )$ for the case of quark--antiquark
interactions, and thus one gets a factor $3$ ($-1$) for the light
quantum numbers of the $\pi$ ($\rho$) meson. In averaged, both
potentials, AL1 and AL1$\chi$, are strongly attractive (moderately
repulsive) for the $\pi$ ($\rho$) quantum numbers, and have been
adjusted to provide the same ground state pion (rho) mass. For
baryons, the AL1 potential gets reduced its strength by a factor of
two (because of the $V_{ij}^{qq} = V_{ij}^{q \bar q}/2$ prescription,
coming from a $\vec{\lambda}_i\vec{\lambda}_j$ color dependence),
however this is not the case for the light sector of the AL1$\chi$
potential. In this latter case, the OGE and confinement terms follow
the former prescription, but while the scalar exchange remains
unaltered, the OPE term is three times more attractive for the
$\Lambda_Q$ baryon than in the pion. The resulting effect is that the
light sector of the AL1$\chi$ interaction in the $\Lambda_Q$ baryon
gets reduced its strength by a factor significantly smaller than 2.
The OPE exchange plays a minor role in the spin triplet
states\footnote{For instance, while the OPE is responsible for around
150 MeV of the binding energy of the pion, it contributes only about 8
MeV to the rho meson mass. } ($\rho$, $\Sigma_Q$ and $\Sigma_Q^*$),
and that explains why the AL1 and AL1$\chi$ potentials predict similar
$\Sigma_Q$ and $\Sigma_Q^*$ masses.

\begin{table}
\begin{tabular}{lc|cccccc|cccccc}
            &&\multicolumn{6}{c}{Charm}&\multicolumn{6}{c}{Bottom} 
\\\tstrut
Baryon      &     & BD    & AL2   & AP1   & AP2   & AL1   & AL1$\chi$ 
                  & BD    & AL2   & AP1   & AP2   & AL1   & AL1$\chi$ 
\\\hline\tstrut
            & FAD & 0.097 & 0.101 & 0.109 & 0.103 & 0.104 &   $-$ 
                  & 0.043 & 0.044 & 0.046 & 0.044 & 0.045 &   $-$    \\
~~~$\Lambda$& VAR & 0.100 & 0.101 & 0.108 & 0.106 & 0.106 &  0.101
                  & 0.045 & 0.043 & 0.048 & 0.045 & 0.045 &  0.042   \\\hline
            & FAD & 0.111 & 0.117 & 0.128 & 0.121 & 0.121 &   $-$ 
                  & 0.051 & 0.053 & 0.056 & 0.053 & 0.054 &   $-$    \\
~~~$\Sigma$ & VAR & 0.112 & 0.118 & 0.127 & 0.118 & 0.123 &  0.123
                  & 0.051 & 0.053 & 0.055 & 0.053 & 0.057 &  0.059   \\\hline
~~~$\Sigma^*$& VAR& 0.122 & 0.131 & 0.138 & 0.131 & 0.135 &  0.139
                  & 0.052 & 0.056 & 0.058 & 0.056 & 0.060 &  0.061   \\\hline
            & FAD & 0.097 & 0.100 & 0.107 & 0.101 & 0.104 &   $-$ 
                  & 0.045 & 0.046 & 0.048 & 0.046 & 0.048 &   $-$    \\
~~~$\Xi$    & VAR & 0.099 & 0.103 & 0.110 & 0.104 & 0.105 &   $-$    
                  & 0.047 & 0.046 & 0.050 & 0.047 & 0.049 &   $-$    \\\hline
~~~$\Xi^\prime$&VAR& 0.106& 0.110 & 0.122 & 0.111 & 0.119 &   $-$   
                  & 0.053 & 0.053 & 0.059 & 0.052 & 0.060 &   $-$    \\\hline
~~~$\Xi^*$  & VAR & 0.115 & 0.123 & 0.133 & 0.126 & 0.123 &   $-$   
                  & 0.052 & 0.057 & 0.060 & 0.054 & 0.059 &   $-$    \\\hline
            & FAD & 0.100 & 0.103 & 0.110 & 0.102 & 0.108 &   $-$ 
                  & 0.050 & 0.052 & 0.055 & 0.051 & 0.054 &   $-$    \\
~~~$\Omega$ & VAR & 0.096 & 0.101 & 0.110 & 0.103 & 0.108 &   $-$    
                  & 0.050 & 0.054 & 0.057 & 0.053 & 0.057 &   $-$    \\\hline
~~~$\Omega^*$&VAR & 0.107 & 0.117 & 0.122 & 0.113 & 0.120 &   $-$    
                  & 0.051 & 0.055 & 0.061 & 0.056 & 0.059 &   $-$    \\\hline
\end{tabular}
\caption{Mass mean square radii in fm$^2$ units, see
Eq.~(\ref{eq:r2m}). Variational and Faddeev~\protect\cite{Si96} 
results are denoted by VAR and FAD,
respectively.}
\label{tab:r2m}
\end{table}
\begin{table}
\begin{tabular}{lc|rrrrrr|rrrrrr}
            &&\multicolumn{6}{c}{Charm}&\multicolumn{6}{c}{Bottom} 
\\\tstrut
Baryon      &     & BD    & AL2   & AP1   & AP2   & AL1   & AL1$\chi$ 
                  & BD    & AL2   & AP1   & AP2   & AL1   & AL1$\chi$ 
\\\hline\tstrut
            & FAD & 0.117 & 0.124 & 0.147 & 0.139 & 0.129 &      
                  & 0.115 & 0.123 & 0.148 & 0.140 & 0.128 &    \\
$\Lambda^+_c$, $\Lambda^0_b$    
            & VAR & 0.120 & 0.125 & 0.145 & 0.142 & 0.131 &  0.124
                  & 0.121 & 0.122 & 0.152 & 0.143 & 0.127 &  0.119       \\\hline
            & FAD & $-$0.224 & $-$0.245 & $-$0.304 & $-$0.287 & $-$0.256 &    
                  & $-$0.280 & $-$0.305 & $-$0.369 & $-$0.347 & $-$0.318 &   \\
$\Sigma^0_c$, $\Sigma^-_b$ 
            & VAR & $-$0.226 &$-$0.248&$-$0.299&$-$0.279&$-$0.261 & $-$0.260
                  & $-$0.279 &$-$0.304&$-$0.357&$-$0.341&$-$0.332 & $-$0.345    \\\hline
            & FAD & 0.134 & 0.145 & 0.174 & 0.164 & 0.151 &    
                  & 0.138 & 0.151 & 0.183 & 0.172 & 0.157 &            \\
$\Sigma^+_c$, $\Sigma^0_b$ 
            & VAR & 0.135 & 0.147 & 0.172 & 0.160 & 0.153 &   0.154
                  & 0.138 & 0.150 & 0.176 & 0.169 & 0.164 &   0.171       \\\hline
            & FAD & 0.494 & 0.535 & 0.652 & 0.615 & 0.557 &    
                  & 0.555 & 0.607 & 0.734 & 0.692 & 0.633 &            \\
$\Sigma^{++}_c$, $\Sigma^+_b$ 
            & VAR & 0.497 & 0.541 & 0.643 & 0.599 & 0.568 &   0.567
                  & 0.555 & 0.603 & 0.710 & 0.679 & 0.659 &   0.687       \\\hline
$\Sigma^{*0}_c$, $\Sigma^{*-}_b$
            & VAR & $-$0.246 & $-$0.274 & $-$0.323 & $-$0.307 & $-$0.283 & $-$0.293
                  & $-$0.289 & $-$0.322 & $-$0.376 & $-$0.364 & $-$0.349 & $-$0.358\\\hline
$\Sigma^{*+}_c$, $\Sigma^{*0}_b$
            & VAR & 0.148 & 0.163 & 0.187 & 0.177 & 0.168 & 0.173        
                  & 0.143 & 0.159 & 0.186 & 0.180 & 0.172 & 0.177   \\\hline
$\Sigma^{*++}_c$, $\Sigma^{*+}_b$
            & VAR & 0.541 & 0.599 & 0.697 & 0.661 & 0.619 & 0.639
                  & 0.574 & 0.639 & 0.747 & 0.725 & 0.694 & 0.711   \\\hline
            & FAD & $-$0.145 & $-$0.154 & $-$0.183 & $-$0.171 & $-$0.161 & 
                  & $-$0.193 & $-$0.203 & $-$0.234 & $-$0.219 & $-$0.212 &  \\
$\Xi^0_c$, $\Xi^-_b$    
           & VAR  & $-$0.148 & $-$0.156 & $-$0.182 & $-$0.174 & $-$0.163 & 
                  & $-$0.198 & $-$0.201 & $-$0.240 & $-$0.220 & $-$0.213 &  \\\hline
           & FAD  & 0.160 & 0.170 & 0.206 & 0.196 & 0.177 &    
                  & 0.151 & 0.161 & 0.197 & 0.187 & 0.168 &  \\
$\Xi^+_c$, $\Xi^0_b$    
           & VAR  & 0.158 & 0.170 & 0.198 & 0.196 & 0.179 &    
                  & 0.150 & 0.151 & 0.189 & 0.180 & 0.158 &  \\\hline
$\Xi^{\prime 0}_c$, $\Xi^{\prime -}_b$      
            & VAR & $-$0.164 & $-$0.179 & $-$0.213 & $-$0.196 &
$-$0.192 &    
                  & $-$0.228 & $-$0.235 & $-$0.285 & $-$0.248 & $-$0.267 &  \\\hline
$\Xi^{\prime +}_c$, $\Xi^{\prime 0}_b$      
            & VAR & 0.174 & 0.199  & 0.234 & 0.218 & 0.207 &    
                  & 0.184 & 0.188  & 0.234 & 0.209 & 0.214 &  \\\hline
$\Xi^{* 0}_c$, $\Xi^{* -}_b$      
            & VAR & $-$0.177 & $-$0.196 & $-$0.232 & $-$0.217 &
$-$0.198 &    
                  & $-$0.224 & $-$0.250 & $-$0.287 & $-$0.259 & $-$0.266 &  \\\hline
$\Xi^{* +}_c$, $\Xi^{* 0}_b$      
            & VAR & 0.190 & 0.218 & 0.254 & 0.241 & 0.213 &    
                  & 0.179 & 0.201 & 0.236 & 0.221 & 0.212 &  \\\hline
            & FAD & $-$0.111 & $-$0.117 & $-$0.131 & $-$0.120 & $-$0.124 &  
                  & $-$0.164 & $-$0.173 & $-$0.191 & $-$0.174 & $-$0.183 &  \\
$\Omega^0_c$, $\Omega^-_b$    
            & VAR & $-$0.108 & $-$0.115 & $-$0.131 & $-$0.120 & $-$0.124 &   
                  & $-$0.163 & $-$0.177 & $-$0.196 & $-$0.180 & $-$0.189 &  \\\hline
$\Omega^{*0}_c$, $\Omega^{*-}_b$  
            &VAR  & $-$0.118 & $-$0.132 & $-$0.143 & $-$0.131 &
$-$0.138 &    
                  & $-$0.167 & $-$0.182 & $-$0.210 & $-$0.189 & $-$0.196 &  \\\hline
\end{tabular}
\caption{Charge mean square radii in $e$fm$^2$ units, see
Eq.~(\ref{eq:r2q}). Variational and Faddeev~\protect\cite{Si96} 
results are denoted by VAR and FAD,
respectively.}
\label{tab:r2q}
\end{table}

\begin{figure}
\vspace{-5cm}
\centerline{\includegraphics[height=30cm]{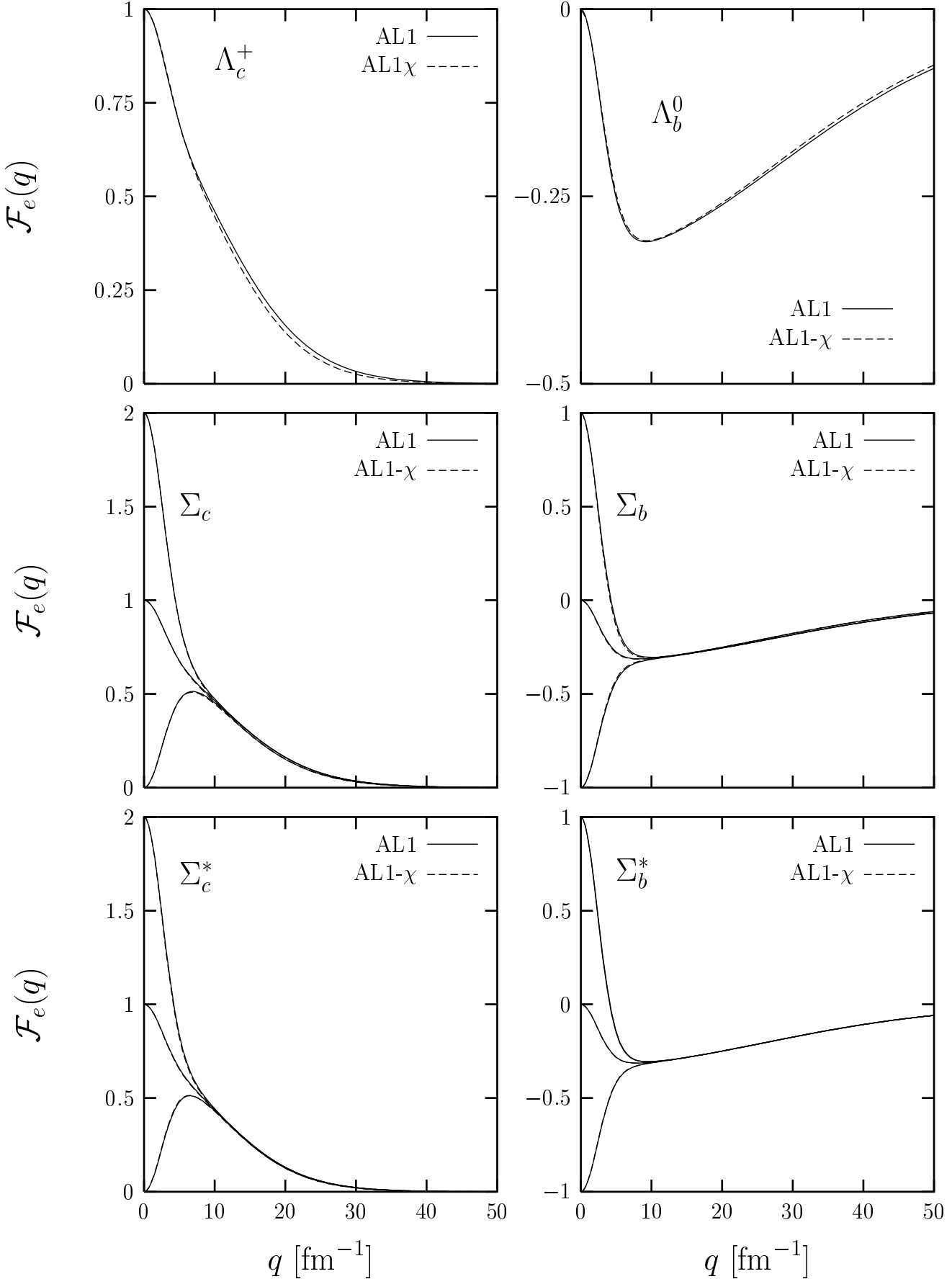}}
\vspace{-1cm}
\caption[pepe]{\footnotesize Charge form factors
(Eq.~(\protect\ref{eq:fe})) of strangeness-less charmed and bottom baryons
obtained with AL1, AL1$\chi$ interactions. The only appreciable
difference, as it is also the case for the spectrum, appears for the
$\Lambda_c$ and $\Lambda_b$ baryons. Note that, ${\cal F}_e(0)$
determines the charge of the baryon.}
\label{fig:fig2}
\end{figure}

\begin{figure}
\vspace{-5cm}
\centerline{\includegraphics[height=30cm]{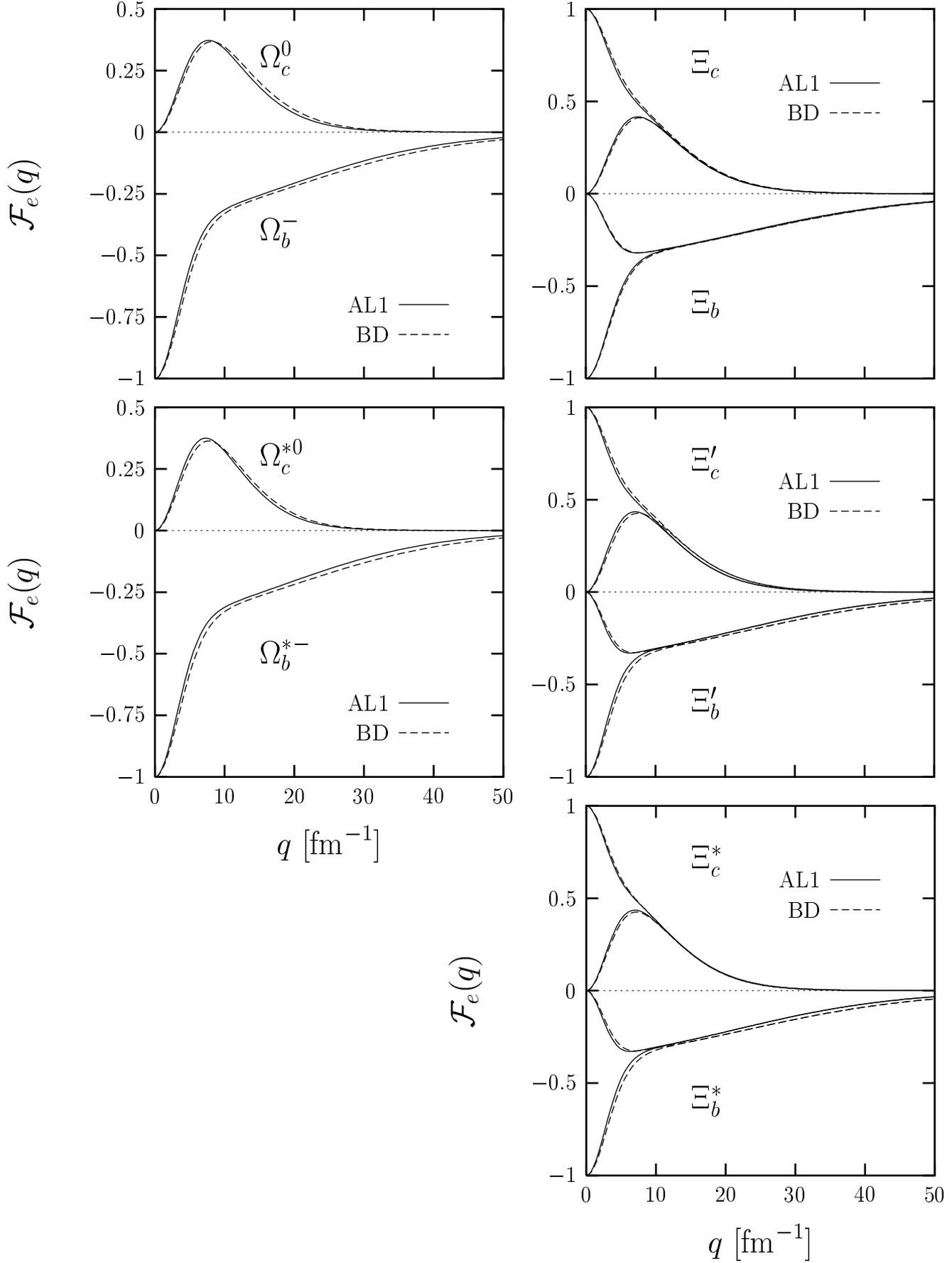}}
\vspace{-1cm}
\caption[pepe]{\footnotesize Charge form factors
(Eq.~(\protect\ref{eq:fe})) of strangeness charmed and bottom baryons
obtained with AL1 and BD quark--quark interactions. The charge of the
baryon is given, in units of the proton charge, by the value of the
form factor at the origin.}
\label{fig:fig3}
\end{figure}

\begin{figure}
\vspace{-5cm}
\centerline{\includegraphics[height=30cm]{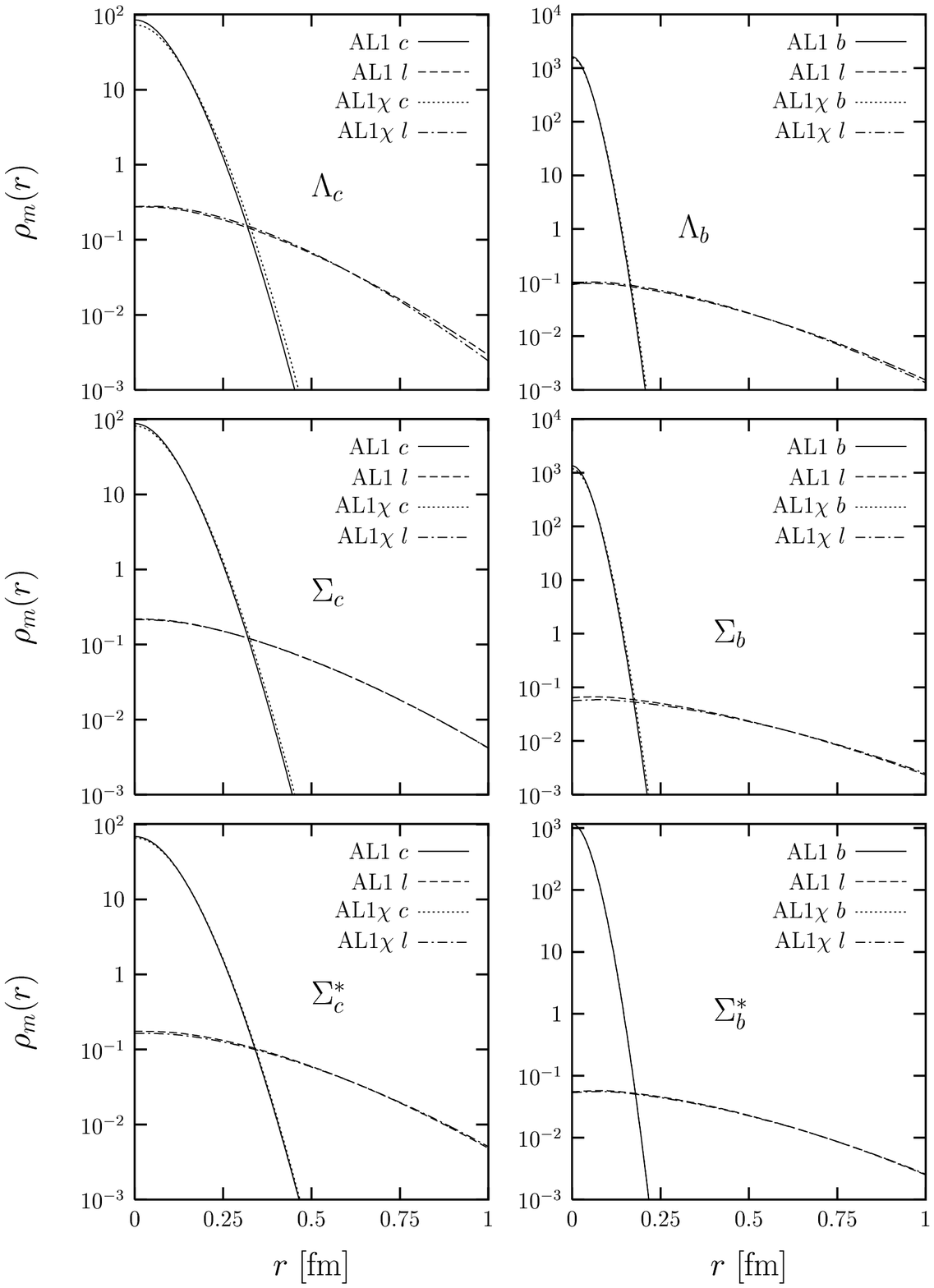}}
\vspace{-1cm}
\caption[pepe]{\footnotesize Quark mass densities [fm$^{-3}$~],
$\rho_m^{B_Q} (\vec{r})\big|_{Q=c,b}$ and $
\rho_m^{B_Q}(\vec{r})\big|_{q=l}$ (Eq.~(\protect\ref{eq:mdens})), of
strangeness-less charmed and bottom baryons obtained with AL1 and
AL1$\chi$ quark--quark interactions. }
\label{fig:fig4}
\end{figure}

\begin{figure}
\vspace{-5cm}
\centerline{\includegraphics[height=28cm]{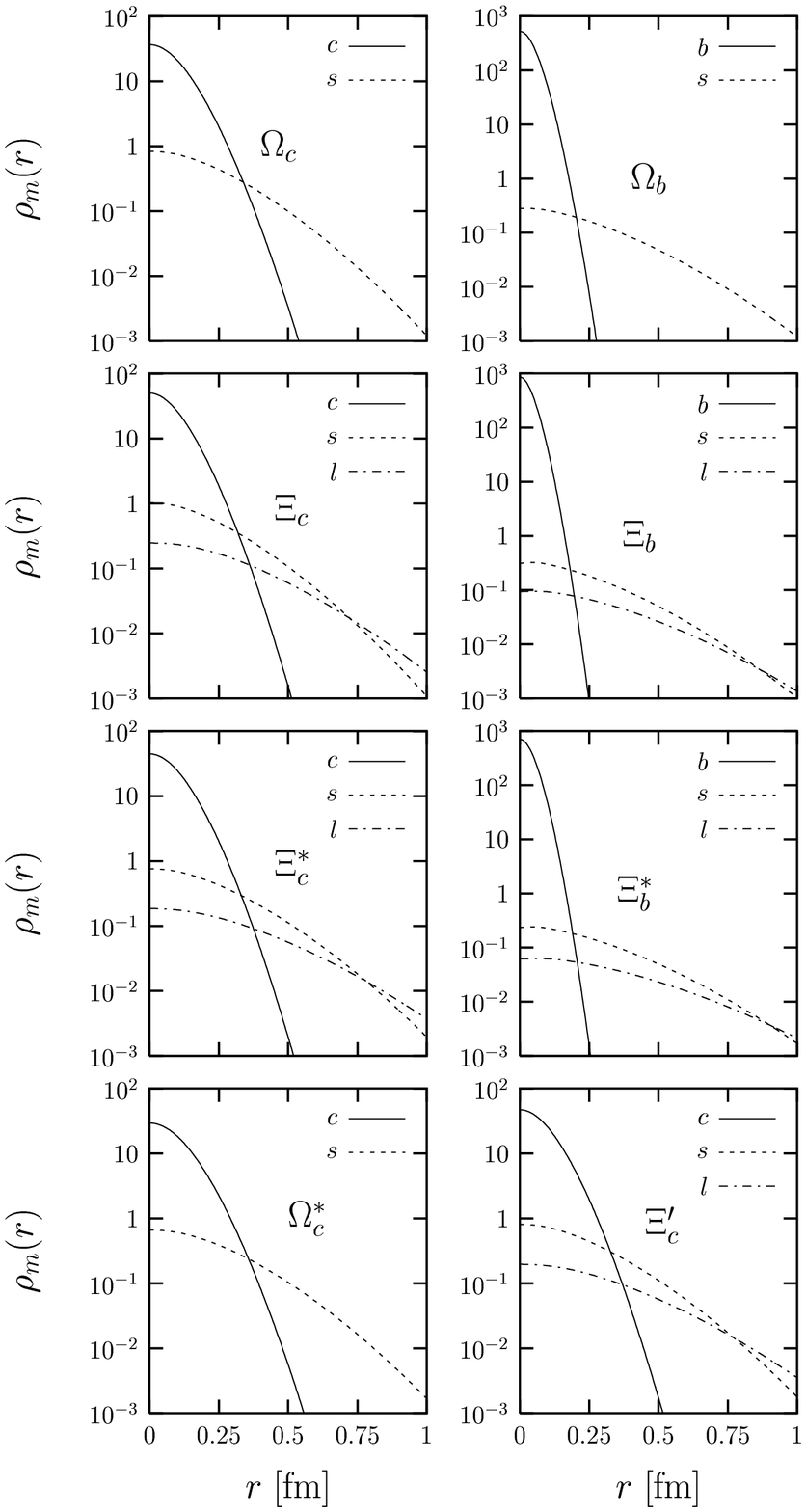}}
\vspace{-1cm}
\caption[pepe]{\footnotesize Quark mass densities [fm$^{-3}$~], $\rho_m^{B_Q}
(\vec{r})\big|_{Q=c,b}$ and $ \rho_m^{B_Q}(\vec{r})\big|_{q=l,s}$
(Eq.~(\protect\ref{eq:mdens})), of  strangeness charmed and bottom baryons
obtained with the AL1 quark--quark interactions.}
\label{fig:fig5}
\end{figure}

We have also computed mass (Table~\ref{tab:r2m}) and charge
(Table~\ref{tab:r2q}) mean square radii, charge form factors 
(Figs.~\ref{fig:fig2}--\ref{fig:fig3}) and quark mass densities (Figs.
\ref{fig:fig4}--\ref{fig:fig5}), for all baryons and the six
inter-quark interactions considered in this work. In
Tables~\ref{tab:r2m} and \ref{tab:r2q}, when possible, we compare our
variational results with those quoted in Ref.~\cite{Si96} and find an
excellent agreement, with tiny differences which at most are of the
order of 3-4\%. Differences between AL1 and AL1$\chi$ potentials
predictions, in Tables~\ref{tab:r2m} and \ref{tab:r2q} and
in Figs.~\ref{fig:fig2}--\ref{fig:fig5}, are also quite small and
it will be quite difficult to use those to experimentally disentangle
among both interactions.

To further test the goodness of our variational baryon wave functions,
we have followed Ref.~\cite{Si96} and we have also computed the so
called wave functions at the origin for the different quark pairs
inside  the heavy baryon (see Eq.~(\ref{eq:eta0})). 
It seems~\cite{Si96} that the
absolute value of this quantity depends dramatically on the numerical
procedure used to solve the three body problem\footnote{Indeed, the
wave function at the origin is especially sensitive to short-range
correlations which are not precisely taken into account by approximate
treatments.} and then is a perfect observable to check the validity of our
approach.
 Results for the
different inter-quark interactions and heavy baryons can be found in
Table~\ref{tab:rho0}, where, when possible, we compare our variational
results with those quoted in Ref.~\cite{Si96}, as well. 
% This observable 
%is of special importance for leptonic and semileptonic
%decays of the heavy baryons,
 For $\Lambda-$, $\Sigma-$ and $\Omega-$type baryons, we
find that our variational results and those obtained from the Faddeev
calculation of Ref.~\cite{Si96}, agree\footnote{Note that for $\Sigma$
baryons and when the BD interaction is used, there exists a serious
discrepancy between Faddeev and variational predictions for the $ll$
pair wave function at the origin. Indeed, the Faddeev prediction is
non-zero, while it is zero within the variational approach. This
is because our variational scheme assumes a pure $S_{\rm light}=1$
configuration for this baryon. It turns out, that the spin triplet
light quark--light quark BD potential is infinitely repulsive for
$r_{12}=0$ and conversely the variational wave function vanishes at
the origin (because of the correlation function $f^{B_Q}$,
in Eq.~(\ref{eq:f12})). However in the Faddeev calculation of
Ref.~\cite{Si96}, the HQS suppressed $S_{\rm light}=0$ component of the
$\Sigma$ wave function is kept, and since for the spin singlet
configuration the BD potential becomes attractive, a non-vanishing
wave function at the origin for the $ll$ pair is found in
Ref.~\cite{Si96}.} within a 10 or 20\%. This constitutes a remarkable
result, since, as it is mentioned in Ref.~\cite{Si96}, it is not
surprising to have approximate treatments of the three body problem
giving values for the wave function at the origin off by almost an
order of magnitude.

However, variational and Faddeev results are in total disagreement for
the $\Xi$ baryons (discrepancies of about a factor three or
larger). We have no explanation for this fact, beyond that saying that
we might have misunderstood the definition of Ref.~\cite{Si96} of the
wave functions at the origin and thus Eq.~(\ref{eq:eta0}) is wrong or
that our $\Xi-$wave functions at the origin are wrong. We very much
doubt this latter possibility, and we would like to point out that the
Faddeev values for the $\Xi-$baryons are abnormally big (about a
factor four), when compared to those quoted for the rest of the
baryons $\Lambda, \Sigma$ and $\Omega$. We are not aware of any reason
to support/explain this fact and indeed we do not observe such a
situation in the variational results.

To finish this section we would like to devote a few words to the
multipolar content of the variational wave functions. In
Tables~\ref{tab:multL} and~\ref{tab:multS} we show the probability
${\cal P}_l$, defined in Eq.~(\ref{eq:mult}), of finding each of the
two light--quarks with angular momentum $l$ and coupled to $L=0$ for
the $\Lambda$ and $\Sigma$ baryons. Results obtained with two
different interactions (AL1 and AL1$\chi$) are compiled, but the gross
features of the multipolar decomposition do not depend significantly
on the selected interaction.  Only multipoles up to the waves $l=3$ or
$l=4$ significantly contribute\footnote{Note that we show probabilities: the
components of the wave function for each value of $l$ are given, up to
a sign, by the squared root of the numbers presented in the tables.}, 
being $l=0$ the dominant one.

\begin{table}
\begin{tabular}{lcc|cccccc|cccccc}
         &          &     &\multicolumn{6}{c}{Charm}&\multicolumn{6}{c}{Bottom} 
\\\tstrut
Baryon        & Pair &     & BD   & AL2  & AP1  & AP2  & AL1  & AL1$\chi$ 
                           & BD   & AL2  & AP1  & AP2  & AL1  & AL1$\chi$ 
\\\hline\tstrut
              & $ll$ & FAD & 23.1 & 15.0 & 12.1 & 11.9 & 14.8 &   $-$ 
                           & 23.3 & 15.2 & 12.2 & 12.1 & 14.9 &   $-$    \\
              & $ll$ & VAR & 22.8 & 18.5 & 14.0 & 13.9 & 18.6 &  54.0 
                           & 19.5 & 19.1 & 14.4 & 14.8 & 18.3 &  74.7     \\
              & $Ql$ & FAD & 8.8  & 7.7  & 6.1  & 6.4  & 7.4  &   $-$ 
                           & 8.8  & 7.7  & 6.1  & 6.4  & 7.4  &   $-$       \\
~~~$\Lambda$  & $Ql$ & VAR & 7.6  & 7.0  & 5.6  & 5.5  & 6.5  &  6.0  
                           & 7.5  & 7.9  & 4.9  & 5.6  & 6.9  &  7.0  \\\hline
              & $ll$ & FAD & 6.9  & 6.6  & 5.0  & 5.6  & 6.1  &   $-$ 
                           & 6.8  & 6.4  & 4.9  & 5.4  & 6.0  &   $-$    \\
              & $ll$ & VAR & 0.0  & 6.8  & 5.2  & 6.2  & 5.7  &  0.0  
                           & 0.0  & 6.9  & 5.5  & 5.3  & 6.7  &  0.0      \\
              & $Ql$ & FAD & 8.2  & 6.7  & 5.3  & 5.5  & 6.5  &   $-$ 
                           & 7.1  & 5.9  & 4.7  & 4.9  & 5.7  &   $-$       \\
~~~$\Sigma$   & $Ql$ & VAR & 7.3  & 6.0  & 4.6  & 5.2  & 5.9  &  5.5  
                           & 6.7  & 6.1  & 3.9  & 4.5  & 4.8  &  4.2  \\\hline
              & $ll$ & VAR & 0.0  & 6.6  & 4.7  & 5.3  & 5.9  &  0.0  
                           & 0.0  & 6.3  & 5.1  & 5.2  & 6.4  &  0.0      \\
~~~$\Sigma^*$ & $Ql$ & VAR & 5.7  & 4.3  & 3.5  & 3.8  & 4.4  &  4.1  
                           & 6.4  & 5.2  & 3.3  & 3.5  & 4.0  &  3.9  \\\hline
              & $ls$ & FAD & 70.5 & 54.6 & 51.0 & 52.1 & 53.5 &   $-$ 
                           & 93.1 & 72.0 & 66.8 & 68.4 & 70.2 &   $-$    \\
              & $ls$ & VAR & 21.1 & 21.9 & 17.0 & 18.1 & 19.3 &   $-$ 
                           & 23.8 & 20.4 & 15.6 & 17.2 & 20.2 &   $-$     \\
              & $Ql$ & FAD & 40.0 & 37.0 & 34.4 & 37.0 & 35.5 &   $-$ 
                           & 51.4 & 48.1 & 44.6 & 48.0 & 46.1 &   $-$       \\
              & $Ql$ & VAR & 11.7 & 9.5  &  8.3 &  8.7 & 9.1  &   $-$ 
                           & 11.6 & 11.9 &  8.7 & 10.2 & 10.0 &   $-$    \\
              & $Qs$ & FAD & 30.8 & 27.7 & 25.3 & 26.4 & 27.1 &   $-$ 
                           & 39.9 & 35.5 & 32.4 & 33.4 & 34.9 &   $-$       \\
~~~$\Xi$      & $Qs$ & VAR & 8.8  & 7.5  &  6.1 & 6.8  & 7.8  &   $-$ 
                           & 8.7  & 8.3  &  6.1 & 7.2  & 7.3  &   $-$ \\\hline
              & $ls$ & VAR & 12.5 & 10.3 &  8.4 & 9.2  & 9.6  &   $-$ 
                           & 11.9 & 10.1 &  8.2 & 9.1  & 9.2  &   $-$     \\
              & $Ql$ & VAR & 11.3 & 9.5  &  7.1 & 9.0  & 8.2  &   $-$ 
                           & 9.8  & 9.5  &  5.8 & 8.6  & 6.7  &   $-$    \\     
~~~$\Xi^\prime$&$Qs$ & VAR & 8.7  & 8.2  &  5.8 & 6.9  & 6.5  &   $-$ 
                           & 8.2  & 7.3  &  4.7 & 6.3  & 5.5  &   $-$ \\\hline
              & $ls$ & VAR & 12.1 & 8.8  &  6.8 & 7.9  & 8.9  &   $-$ 
                           & 12.1 & 9.3  &  7.9 & 8.4  & 9.4  &   $-$     \\
              & $Ql$ & VAR & 8.5  & 6.4  &  5.8 & 6.0  & 7.7  &   $-$ 
                           & 10.4 & 8.0  &  5.7 & 7.7  & 6.8  &   $-$    \\     
~~~$\Xi^*$    & $Qs$ & VAR & 6.7  & 5.7  &  4.8 & 4.6  & 6.1  &   $-$ 
                           & 8.4  & 6.3  &  4.7 & 6.0  & 5.5  &   $-$ \\\hline
              & $ss$ & FAD & 16.5 & 15.5 & 13.8 & 15.3 & 14.6 &   $-$ 
                           & 16.7 & 15.6 & 13.7 & 15.3 & 14.6 &   $-$    \\
              & $ss$ & VAR & 17.7 & 17.2 & 14.8 & 16.9 & 15.7 &   $-$ 
                           & 19.9 & 15.8 & 13.1 & 16.9 & 15.5 &   $-$     \\
              & $Qs$ & FAD & 17.8 & 15.4 & 14.4 & 15.0 & 15.1 &   $-$ 
                           & 16.0 & 13.8 & 12.8 & 13.3 & 13.5 &   $-$       \\
~~~$\Omega$   & $Qs$ & VAR & 17.0 & 14.5 & 12.3 & 13.5 & 13.0 &   $-$    
                           & 15.3 & 12.5 & 10.8 & 11.6 & 11.1 &   $-$   \\\hline
              & $ss$ & VAR & 18.7 & 13.5 & 13.5 & 14.9 & 13.0 &   $-$ 
                           & 18.4 & 15.7 & 12.5 & 15.0 & 14.3 &   $-$     \\
~~~$\Omega^*$ & $Qs$ & VAR & 12.6 & 9.9  &  9.4 & 10.2 & 9.7  &   $-$    
                           & 15.0 & 11.7 &  9.4 & 10.7 & 10.4 &   $-$   \\\hline
\end{tabular}
\caption{ Wave function at the origin, $\eta_i^{B_Q}(0)$ in fm$^{-3}$
units, see Eq.~(\ref{eq:eta0}). In the second column, $ll$ and $ss$
pairs stand for $\eta_Q^{B^{ll}_Q}(0)$ and $\eta_Q^{B^{ss}_Q}(0)$.
Besides for strange [strangeness-less ] baryons, the pairs $Ql$ and
$Qs$ stand for $\eta_{q^\prime=s}^{B^{ls}_Q}(0) $ $\left
[\eta_{q=l}^{B^{ll}_Q}(0) = \eta_{q^\prime=l}^{B^{ll}_Q}(0) \right ]$
and $\eta_{q=l}^{B^{ls}_Q}(0)$.  Variational and
Faddeev~\protect\cite{Si96} results are denoted by VAR and FAD,
respectively.  For the $\Sigma$ and $\Sigma^*$ baryons, our
variational scheme assumes a pure $S_{\rm light}=1$ configuration. 
The VAR estimates for $\eta_Q^{B^{ll}_Q}(0)$, and BD and AL1$\chi$
interactions,  vanish for these baryons, 
since  the spin triplet light quark--light
quark forces are infinitely repulsive 
at $r_{12}=0$, for these potentials.}\label{tab:rho0}
\end{table}
\begin{table}
\begin{center}
\begin{tabular}{c|ccccccc|ccccccc}
&\multicolumn{7}{c}{Charm}&\multicolumn{7}{c}{Bottom} 
\\\tstrut
           $l$ &  0 & 1 & 2 & 3 & 4 & 5 & 6 
                         & 0 & 1 & 2 & 3 & 4 & 5 & 6 \\\hline\tstrut
AL1$\chi$--${\cal P}_l$ 
                         & 0.75495 & 0.18706 & 0.04066 & 0.01103 & 0.00368 & 0.00143  & 0.00062
                         & 0.79079 & 0.16186 & 0.03237 & 0.00894 & 0.00324 & 0.00139  & 0.00067\\\tstrut
 $\sum_{k=0}^{l}{\cal P}_k$
                         & 0.7550 & 0.9420 & 0.9827 & 0.9937 & 0.9974 & 0.9988  & 0.9994                    
                         & 0.7908 & 0.9526 & 0.9850 & 0.9940 & 0.9972 & 0.9986  & 0.9993 \\\hline\tstrut
  AL1--${\cal P}_l$      
                         & 0.87389 & 0.10972 & 0.01346 & 0.00227 & 0.00052  & 0.00015  & 0.00005
                         & 0.87368 & 0.11207 & 0.01221 & 0.00169 & 0.00032  & 0.00008  & 0.00003\\\tstrut
 $\sum_{k=0}^{l}{\cal P}_k$ 
                         & 0.8739 & 0.9836 & 0.9971 & 0.9993 & 0.9999 & 1.0000  & 1.0000
                         & 0.8737 & 0.9857 & 0.9980 & 0.9996 & 1.0000 & 1.0000  & 1.0000 \\\hline
\end{tabular}
\end{center}
\caption{Probabilities ${\cal P}_l$ 
(defined in Eq.~(\protect\ref{eq:mult})) for several waves. Results
were obtained for the $\Lambda$ baryon and AL1 and AL1$\chi$
inter-quark interaction. The errors
are always less than one unit in the last digit.  }
\label{tab:multL}
\end{table} 
\begin{table}
\begin{center}
\begin{tabular}{c|rrrrrrr|rrrrrrr}
&\multicolumn{7}{c}{Charm}&\multicolumn{7}{c}{Bottom} 
\\\tstrut
           $l$ &  0 & 1 & 2 & 3 & 4 & 5 & 6 
                         & 0 & 1 & 2 & 3 & 4 & 5 & 6 \\\hline\tstrut
AL1$\chi$--${\cal P}_l$
                        & 0.95074 & 0.04711 & 0.00210 & 0.00014 & 0.00002 & 0.00000  & 0.00000 
                        & 0.95010 & 0.04777 & 0.00204 & 0.00013 & 0.00002 & 0.00000  & 0.00000 \\\tstrut
 $\sum_{k=0}^{l}{\cal P}_k$ 
                        & 0.9507  & 0.9978  & 0.9999  & 1.0000  & 1.0000 & 1.0000  & 1.0000
                        & 0.9501  & 0.9979  & 0.9999  & 1.0000  & 1.0000 & 1.0000  & 1.0000  \\\hline\tstrut
  AL1--${\cal P}_l$     
                        & 0.96542 & 0.03377 & 0.00091 & 0.00003 & 0.00000 & 0.00000  & 0.00000
                        & 0.96109 & 0.03734 & 0.00153 & 0.00011 & 0.00001 & 0.00000  & 0.00000 \\\tstrut
 $\sum_{k=0}^{l}{\cal P}_k$
                        & 0.9654  & 0.9992  & 1.0000  & 1.0000  & 1.0000 & 1.0000  & 1.0000
                        & 0.9611  & 0.9984  & 1.0000  & 1.0000  & 1.0000 & 1.0000  & 1.0000 \\\hline
\end{tabular}
\end{center}
\caption{ Same as Table~\protect\ref{tab:multL} for 
the $\Sigma$ baryon.}
\label{tab:multS}
\end{table}


\section{Conclusions and Outlook}\label{sec:concl}

We have developed a variational scheme to describe the charm and
bottom baryons compiled in Table~\ref{tab:summ}. We compute baryon
masses and wave functions from the variational principle applied to a
family of Jastrow type functions constrained by HQS. For several
inter-quark interactions and baryons, we reproduce the Faddeev results
of Ref.~\cite{Si96} for masses, charge and mass radii, and wave
functions at the origin. Thus, we conclude that the finite heavy quark
mass corrections to the wave-functions should be small, even for
charmed baryons. Besides of giving results for baryons not studied in
Ref.~\cite{Si96} ($\Sigma^*,\Xi^\prime, \Xi^*$ and $\Omega^*$), we
provide, in Tables~\ref{tab:charmparam} and \ref{tab:bottomparam} in
the Appendix, wave functions, parameterized in a simple manner, for
all baryons compiled in Table~\ref{tab:summ} and six different
inter-quark interactions.  Thanks to HQS, the baryon wave functions
are significantly simpler and more manageable than those obtained from
Faddeev calculations and they can be used by the community to compute
further observables, without having to solve the involved three body
problem.  Hence, we have shown in this context, once more, that HQS is
a useful tool to understand the bottom and charm physics.  We have
also worked out the predictions for the strangeness-less baryons of
the SU(2) chirally inspired quark-quark interactions developed in
Ref.~\cite{BFV99,Fe93}. The chirally inspired potential, AL1$\chi$,
predicts masses for the $\Lambda_Q$ baryons of about 155 (charm
sector) and 168 (bottom sector) MeV smaller than those obtained from
the phenomenological AL1 interaction. Assuming that the inclusion of a
pattern of SCSB should lead to a more theoretically founded scenario
to address hadron spectroscopy, the findings of this work might hint
the existence of sizeable three quarks forces with the $\Lambda_Q$
quantum numbers. New SU(3) chirally inspired quark-quark interactions,
valid also for the $s$ quark (exchanges of eta and kaon mesons are
also introduced) are being developed by the Salamanca
group~\cite{ff}. It would be interesting to work out the predictions
for the strangeness baryons with this new set of chirally inspired
interactions, once the potential becomes available.

Using the semi--analytical wave functions found in this work, we plan
to compute the semileptonic decay of the bottom baryons into the
charmed ones. The mixing parameter $|V_{cb}|$ of the
Cabibbo-Kobayashi-Maskawa matrix can be determined from the study of
these processes. Furthermore, the Isgur-Wise form-factor, which is a
universal function that governs all the exclusive $b \to c$
semileptonic decays in the $m_Q\to \infty$ limit, could be obtained
and more importantly the size of the $1/m_Q$ corrections in these
decays would be also estimated.  We will compare our results to those
obtained in lattice QCD simulations, both from the meson and baryon
sectors~\cite{3pt}.

\begin{acknowledgments}
This research was supported by DGI and FEDER funds, under contracts
BFM2002-03218 and BFM2003-00856 and by the Junta de Andaluc\'\i a and
Junta de Castilla y Le\'on under contracts FQM0225 and
SA104/04. C. Albertus wishes to acknowledge a grant related to his
Ph.D from Junta de Andaluc\'\i a.

\end{acknowledgments}

\section{Appendix: variational wave function parameters}

All VAR results presented in the paper have been obtained with
light-light correlation functions, $F^{B_Q}$, constructed from a
linear combination of four gaussians (see eq.~(\ref{eq:fbig12})). In
Table~\ref{tab:ng} and for the $\Lambda_c$ baryon with an AL1
inter-quark interaction, we present the dependence of the mass and the
value of the wave function at the origin on the number of gaussians
used to build up the correlation function.
\begin{table}
\begin{tabular}{c|c|c|c}
Number of gaussians & $\Lambda_c$ Mass [MeV]  & $\eta^{\Lambda_c}_{ll}(0)$
[fm$^{-3}$] &  $\eta^{\Lambda_c}_{cl}(0)$ [fm$^{-3}$] \\\hline\tstrut
1 & 2302.1  & 16.4 & 5.2 \\
2 & 2295.0  & 17.0 & 5.7 \\
3 & 2294.8  & 18.6 & 6.2 \\
4 & 2294.6  & 18.6 & 6.5 \\\hline\tstrut
FAD & 2296  & 14.8 & 7.4 \\\hline
\end{tabular}
\caption{$\Lambda_c$ baryon mass and wave functions at the origin (see
Eq.~(\ref{eq:eta0})) as a function of the number of gaussians included
in the light--light correlation function. Results have been obtained
with the AL1 inter--quark interaction  and
for comparison, the Faddeev results of Ref.\protect\cite{Si96}
(denoted by FAD) have been also compiled.  }\label{tab:ng}
\end{table}
Finally, we present in Tables~\ref{tab:charmparam} and
\ref{tab:bottomparam} the variational parameters of the charmed and
bottom baryon wave functions for the different quark-quark
interactions analyzed in this work. The wave functions are easy to
handle and can be used to evaluate further observables.\vspace{1cm}

\begin{table}[h]
\begin{tabular}{ll|rrrrrrrrrrrrr}
             &&\multicolumn{12}{c}{Charm} \\\tstrut
             &         &$b_1$ &$d_1$ &~$a_2$   &$b_2$ &  $d_2$&~$a_3$   &$b_3$ &$d_3$ &~$a_4$   &$b_4$ &$d_4$ &~$\alpha_q$&~$\alpha_{q^\prime}$\\\hline\tstrut
             &AL1      & 0.51 & 0.76 &  0.67   & 0.73 &  0.73 &    0.56 & 0.85 & 0.76 &   \phantom{$-$}0.77 & 1.12 & 0.84 &   \phantom{$-$} 0.19   &    0.19 \\
             &AL1$\chi$& 0.57 & 1.51 &  0.68   & 1.11  & 1.02  & 1.31    &1.49  & 1.10 & 0.82    &1.27  & 0.95 & 0.39      & 0.39    \\
             &AL2      & 1.07 & 0.37 &  0.62   & 0.65 & 0.93  & 0.51    & 0.99 & 0.49 &    1.13 & 0.45 & 0.94 &  0.18     &  0.18   \\
             &AP1      & 0.56 & 0.64 &  0.69   & 0.53 & 0.62  & 0.64    & 0.79 & 0.68 &    0.92 & 0.92 & 0.76 &  0.13     &  0.13   \\
             &AP2      & 0.72 & 0.77 &  0.79   & 0.72 & 0.50  & 0.78    & 0.46 & 0.65 &    0.72 & 0.81 & 0.77 &  0.20     &  0.20   \\
$\Lambda$    &BD       & 1.08 & 0.86 &  0.34   & 0.45 & 0.61  & 0.79    & 0.56 & 0.83 &    1.28 & 0.87 & 0.74 &  0.22     &  0.22   \\\hline
             &AL1      & 0.55 &$-$0.12 & $-$0.61 & 0.62 &  0.01 &    0.52 & 0.63 & 0.24 &    0.77 & 0.50 & 0.27 &    0.06   &    0.06 \\
             &AL1$\chi$& 0.46 & 0.44 &  0.63   & 0.51 & 0.50  & 0.54    & 0.50 & 0.45 &    0.61 & 0.54 & 0.62 & 0.15      &  0.15   \\
             &AL2      & 0.71 &$-$0.12 &1.00     & 0.56 & 0.50  & 0.51    & 0.44 & 0.26 &    1.17 & 0.43 & 0.68 & 0.13      & 0.13    \\
             &AP1      & 0.43 & 0.42 & 0.75    & 0.51 & 0.39  & 0.53    &0.38  & 0.37 &    0.62 & 0.49 & 0.48 & 0.10      &  0.10   \\
             &AP2      & 0.42 & 0.43 & 0.80    & 0.42 & 0.47  & 0.54    &0.52  & 0.38 &    0.68 & 0.55 & 0.70 & 0.07      &  0.07   \\
$\Sigma$     &BD       & 0.58 & 0.90 & 0.99    & 0.48 & 0.67  & 0.76    &0.33  & 0.71 &    1.08 & 0.54 & 0.65 & 0.08      &  0.08   \\\hline
             &AL1      & 0.49 & 0.46 &    0.61 & 0.50 &  0.45 &    0.63 & 0.53 & 0.40 &    0.69 & 0.43 & 0.52 &    0.33   &    0.33 \\
             &AL1$\chi$& 0.46 & 0.45 & 0.62    & 0.45 &  0.46 &  0.69   & 0.54 & 0.38 &    0.72 & 0.43 & 0.57 &  0.37     & 0.37    \\
             &AL2      & 0.67 & 0.44 & 0.92    & 0.49 &  0.70 &  0.86   & 0.53 & 0.05 &    0.90 & 0.44 & 0.48 &  0.41     & 0.41    \\
             &AP1      & 0.42 & 0.43 & 0.64    & 0.48 &  0.42 &  0.64   & 0.46 &  0.38&    0.69 & 0.49 & 0.53 &  0.32     & 0.32    \\
             &AP2      & 0.45 & 0.51 & 0.63    & 0.50 &  0.47 &  0.64   & 0.51 & 0.39 &    0.62 & 0.41 & 0.53 &  0.32     & 0.32    \\
$\Sigma^*$   &BD       & 0.75 & 0.87 & 0.93    & 0.45 &  0.68 &  0.75   & 0.51 & 0.58 &    1.22 & 0.41 & 0.65 &  0.27     & 0.27    \\\hline
             &AL1      & 0.65 & 0.86 &    0.35 & 0.73 &  0.56 &    0.69 & 0.50 & 0.81 &    0.68 & 1.06 & 0.62 &   0.34 & 0.15 \\
             &AL2      & 0.78 & 0.49 &    0.82 & 0.52 &  0.47 &  1.06   & 0.69 & 0.73 &    1.65 & 1.12 & 0.62 &   0.37    &   0.25  \\
             &AP1      & 1.10 & 0.45 &    0.09 & 0.63 &  0.47 &  0.55   & 0.74 &0.51  &    0.65 & 0.55 & 0.28 &   0.28    &   0.25  \\
             &AP2      & 0.92 & 0.71 &    0.01 & 0.69 &  0.49 &  0.51   & 0.87 & 0.37 &    0.54 & 0.48 & 0.43 &   0.32    &   0.20  \\
$\Xi$        &BD       & 0.82 & 0.82 &   0.45  & 0.33 &  0.57 &  0.77   & 0.60 & 0.75 &    2.10 & 0.86 & 0.55 &   0.23    &   0.18  \\\hline
             &AL1      & 0.53 & 0.51 &    0.64 & 0.58 &  0.54 &
             0.60 & 0.60 & 0.43 &    0.71 & 0.51 & 0.59 &     0.24 &  0.22 \\
             &AL2      & 0.61 & 0.46 &    1.00 & 0.67 & 0.47  &  0.77   & 0.50 & 0.39 &    0.99 & 0.48 & 0.41 &   0.16    &   0.04  \\
             &AP1      & 0.50 & 0.51 &    0.66 & 0.57 & 0.52  &
             0.60 & 0.60 & 0.43 &    0.69 & 0.50 & 0.56 &     0.23   &   0.20  \\
             &AP2      & 0.61 & 0.45 &    0.44 & 0.47 & 0.39  &
             0.74 & 0.42 & 0.49 &    0.69 & 0.57 & 0.52 &    0.08   &   0.08  \\
$\Xi^\prime$  &BD       & 0.58 & 0.78 &    0.75 & 0.53 & 0.61  &    1.01 & 0.52 & 0.62 &    1.35 & 0.59 & 0.61 &    0.12   &   0.11  \\\hline
             &AL1      & 0.55 & 0.47 &    0.60 & 0.55 &  0.47 &    0.59 & 0.54 & 0.46 &    0.66 & 0.53 & 0.57 & 0.28   & 0.28 \\
             &AL2      & 0.53 & 0.18 &    0.71 & 0.82 &  0.31 &  0.71   & 0.59 & 0.37 &    1.36 & 0.69 & 0.16 &   0.54    &   0.40  \\
             &AP1      & 0.49 &  0.47&    0.57 & 0.53 &  0.44 &  0.62   & 0.54 & 0.38 &    0.64 & 0.48 & 0.50 &    0.35   &   0.36  \\
             &AP2      & 0.74 &  0.38&  0.21   & 0.57 & 0.19  & 0.68    &0.57  & 0.21 &    0.64 & 0.47 & 0.20 &    0.44   &    0.50 \\
$\Xi^*$      &BD       & 0.61 &  0.83&   0.69  & 0.49 & 0.69  & 1.09    & 0.64 & 0.74 &    1.48 & 0.52 & 0.67 &    0.39   &   0.36  \\\hline
             &AL1      & 0.66 & 0.43 &    0.10 & 0.59 &  0.58 &    0.92 & 0.56 & 0.55 &    0.75 & 0.71 & 0.65 &    0.17   &    0.17 \\
             &AL2      & 0.71 & 0.46 &  0.86   & 0.75 &  0.43 & 0.80    & 0.55 & 0.42 &    0.83 & 0.69 & 0.50 &   0.13    &   0.13  \\
             &AP1      & 0.78 & 0.14 &    0.97 & 0.50 &  0.97 & $-$0.04  & 0.57 & 0.80 &    0.83 & 1.16 & 1.33 &   0.19    &   0.19  \\
             &AP2      & 0.63 & 0.50 &   0.58  & 0.60 & 0.60  & 0.52    & 0.67 & 0.53 &    0.76 & 0.65 & 0.64 &   0.19    &   0.19  \\
$\Omega$     &BD       & 1.02 & 0.20 &   0.75  & 0.51 & 0.22  & 0.68    & 0.58 & 0.33 &    1.37 & 0.90 & 0.21 &   0.04    &   0.04  \\\hline
             &AL1      & 0.74 & 0.49 &    0.46 & 0.59 &  0.49 &    0.63 & 0.66 & 0.38 &    0.74 & 0.52 & 0.45 &    0.44   &    0.44 \\
             &AL2      & 0.86 & 0.17 &   0.43  & 0.50 &  0.47 & 0.38    & 0.48 & 0.53 &    0.70 & 0.67 & 0.62 &    0.50   &  0.50   \\
             &AP1      & 0.70 & 0.61 &   0.40  & 0.61 &  0.54 &  0.75   & 0.60 & 0.49 &    0.72 & 0.54 & 0.60 &    0.46   &  0.46   \\
             &AP2      & 0.67 & 0.49 &   0.25  & 0.77 & 0.61  & 0.55    & 0.74 & 0.45 &    0.69 & 0.56 & 0.43 &    0.47   & 0.47    \\
$\Omega^*$   &BD       & 0.46 & 1.10 &   0.64  & 0.61 & 0.77  & 1.15    & 0.73 & 0.75 &    1.32 & 0.63 & 0.77 &    0.34   & 0.34    \\\hline
\end{tabular}
\caption{Variational parameters ($a's$ are dimensionless, $d's$ have
dimensions of fm and $b's$ have dimensions of fm$^{-1}$) of the baryon
three body wave function (Eqs.~(\protect\ref{eq:onebody})
and~(\protect\ref{eq:fbig12})), in the charm sector and for different
inter-quark interactions. Besides for the BD and AL1$\chi$
interactions, the $c$  parameter
(Eq.~(\protect\ref{eq:f12})) has been set to 200 fm$^{-1}$ for the
 $\Sigma $ and $\Sigma^*$  baryons. 
In the rest of cases, $c$ is set to $+\infty$.}
\label{tab:charmparam}
\end{table}
%
%
%\vspace{2cm}
\begin{table}[h!]
\begin{tabular}{ll|rrrrrrrrrrrrr}
             &&\multicolumn{12}{c}{Bottom} \\\tstrut
             &         &$b_1$ &$d_1$ &~$a_2$   &$b_2$ &  $d_2$&~$a_3$   &$b_3$ &$d_3$ &~$a_4$   &$b_4$ &$d_4$ &$\alpha_q$&~$\alpha_{q^\prime}$\\\hline\tstrut
             &AL1      & 0.60 & 0.54 &    \phantom{$-$}0.58 & 0.71 &  0.76 &    \phantom{$-$}0.46 & 0.75 & 0.72 &    \phantom{$-$}0.78 & 0.84 & 0.77 &    0.00   &    0.00 \\       
             &AL1$\chi$& 0.48 & 2.67 & 1.09    & 1.67 & 1.40  & 1.74    &1.54  & 1.04 &    0.66 & 1.30 & 1.59 &    0.09   &    0.09 \\
             &AL2      & 0.45 & 0.29 &  1.21   & 0.63 & 0.91  & 0.96    & 0.97 & 0.62 &    1.73 & 0.88 & 0.63 &$-$0.08     &$-$0.08    \\
             &AP1      & 0.56 & 0.67 &  0.72   & 0.53 & 0.65  & 0.66    & 0.80 & 0.70 &    0.94 & 0.91 & 0.78 &  0.09     &  0.09   \\
             &AP2      & 0.96 & 0.68 &  0.41   & 0.70 & 0.47  & 0.55    & 0.72 & 0.42 &    0.55 & 0.43 & 0.58 &  0.04     &  0.04   \\
$\Lambda$    &BD       & 0.78 & 0.95 &  0.99   & 0.43 & 0.76  & 1.22    & 0.73 & 0.79 &    1.24 & 0.77 & 0.82 &  0.04     &  0.04   \\\hline
             &AL1      & 0.38 & 0.18 &    1.17 & 0.48 &  0.44 &    0.69 & 0.58 & 0.33 &    0.86 & 0.59 & 0.74 &    0.08   &    0.08 \\       
             &AL1$\chi$& 0.49 & 0.44 &  0.68   & 0.47 & 0.50  & 0.59    & 0.54 & 0.38 &    0.65 & 0.46 & 0.67 &   0.21    &  0.21   \\
             &AL2      & 0.65 & 0.06 &  0.95   & 0.56 & 0.71  & 0.29    & 0.67 & 0.19 &    1.31 & 0.22 & 0.58 &$-$0.08      &$-$0.08    \\
             &AP1      & 0.43 & 0.39 &  1.09   & 0.51 & 0.44  & 0.99    & 0.39 & 0.41 &    1.09 & 0.54 & 0.55 &   0.09    &  0.09   \\
             &AP2      & 0.40 & 0.46 &  0.71   & 0.45 & 0.41  & 0.71    & 0.50 & 0.29 &    0.75 & 0.37 & 0.53 &   0.02    &  0.02   \\
$\Sigma$     &BD       & 0.64 & 0.50 &  1.07   & 0.58 & 0.40  & 1.71    & 0.47 & 0.35 &    1.77 & 0.37 & 0.54 &$-$0.05    &$-$0.05  \\\hline
             &AL1      & 0.44 & 0.45 &    0.63 & 0.52 &  0.48 &    0.58 & 0.49 & 0.41 &    0.68 & 0.54 & 0.57 &    0.23   &    0.23 \\      
             &AL1$\chi$& 0.39 & 0.51 & 0.69    & 0.55 & 0.40  &  0.60   & 0.46 & 0.41 &    0.60 & 0.52 & 0.45 &   0.23    &   0.23  \\
             &AL2      & 0.56 & 0.12 & 0.98    & 0.45 & 0.72  &  0.48   & 0.55 & 0.31 &    1.14 & 0.32 & 0.61 &  0.04     & 0.04    \\
             &AP1      & 0.39 & 0.51 &  0.70   & 0.55 & 0.40  &   0.59  & 0.46 & 0.41 &    0.60 & 0.52 & 0.45 &  0.22     & 0.22    \\
             &AP2      & 0.38 & 0.50 &  0.70   & 0.54 & 0.39  &  0.60   & 0.45 & 0.41 &    0.59 & 0.52 & 0.45 &  0.23     & 0.23    \\
$\Sigma^*$   &BD       & 0.71 & 0.74 &  0.87   & 0.31 & 0.62  &  0.83   & 0.40 & 0.60 &    1.30 & 0.44 & 0.65 &$-$0.014     &$-$0.014   \\\hline
             &AL1      & 0.62 & 0.61 &    0.60 & 0.75 &  0.72 &    0.54 & 0.82 & 0.62 &    0.76 & 0.71 & 0.70 & 0.10   & 0.06 \\
             &AL2      & 0.87 & 0.40 & 0.97    & 0.63 & 0.58  & 0.69    & 0.48 & 0.56 &    0.93 & 0.97 & 0.73 &$-$0.03     &$-$0.03   \\
             &AP1      & 0.92 &  0.63& 0.47    &0.64  &  0.50 &  0.71   &0.74  & 0.40 &    0.68 & 0.46 & 0.54 &   0.06   &    0.04\\
             &AP2      & 0.95 & 0.74 & 0.50    & 0.68 &  0.48 & 0.81    &0.71  & 0.40 &    0.59 & 0.43 & 0.56 &   0.00   &  $-$0.05 \\
$\Xi$        &BD       & 0.84 & 0.85 & 1.13    & 0.52 &  0.86 & 0.18    &0.73  & 1.06 &    1.10 & 1.46 & 1.12 &   0.06   &   0.01 \\\hline
             &AL1      & 0.50 & 0.51 &    0.66 & 0.57 &  0.52 &    0.60 & 0.60 & 0.43 &    0.69 & 0.50 & 0.56 & 0.23   & 0.20 \\
             &AL2      & 0.69 & 0.21 &  1.12   & 0.53 &  0.42 & 0.58    & 0.41 &  0.21&    1.29 & 0.48 & 0.68 &  $-$0.03   & $-$0.03  \\
             &AP1      & 0.49 & 0.51 &   0.67  & 0.54 &  0.52 & 0.63    & 0.57 & 0.46 &    0.69 & 0.51 & 0.59 &   0.23   &   0.20 \\
             &AP2      & 0.49 & 0.48 &   0.57  & 0.52 &  0.43 & 0.64    & 0.53 & 0.38 &    0.64 & 0.47 & 0.49 & $-$0.04    &$-$0.03   \\
$\Xi^\prime$  &BD       & 0.74 & 0.92 &   0.90  & 0.33 & 0.62  & 0.89    & 0.50 & 0.63 &    1.26 & 0.49 & 0.72 &  0.02    & $-$0.05  \\\hline
             &AL1      & 0.50 & 0.51 &    0.66 & 0.57 &  0.52 &    0.60 & 0.60 & 0.51 &    0.69 & 0.50 & 0.56 & 0.22   & 0.20 \\
             &AL2      & 0.85 & 0.06 &  1.00   & 0.59 &  0.50 & 0.48    &0.44  & 0.26 &    0.87 & 0.33 & 0.67 &  0.11    &  0.11  \\
             &AP1      & 0.49 & 0.52 &    0.68 & 0.51 &  0.51 &  0.65   & 0.53 & 0.48 &    0.70 & 0.52 & 0.62 &  0.23    &  0.20  \\
             &AP2      & 0.57 & 0.48 &    0.62 & 0.40 &  0.39 &  0.80   & 0.44 & 0.40 &    0.74 & 0.48 & 0.58 &  0.02    &  0.00  \\
$\Xi^*$      &BD       & 0.77 & 0.86 &    0.90 & 0.33 &  0.63 & 0.86    & 0.46 & 0.63 &    1.28 & 0.50 & 0.71 & $-$0.03    & $-$0.07  \\\hline
             &AL1      & 0.60 & 0.45 &    0.58 & 0.61 &  0.58 &    0.49 & 0.55 & 0.50 &    0.72 & 0.66 & 0.66 &    0.13   &    0.13 \\
             &AL2      & 0.62 & 0.31 &    0.67 & 0.65 & 0.56  & 0.50    & 0.59 & 0.32 &    0.88 & 0.65 & 0.47 &  0.06     &  0.06   \\
             &AP1      & 0.72 & 0.26 &    0.69 & 0.66 &  0.32 &   0.51  & 0.62 & 0.22 &    0.75 & 0.57 & 0.26 &   0.08    &  0.08   \\
             &AP2      & 0.63 & 0.45 &    0.78 & 0.62 &  0.58 &   0.49  & 0.55 & 0.50 &    0.72 & 0.66 & 0.66 &   0.13    &  0.13   \\
$\Omega$     &BD       & 0.82 & 0.85 &    0.89 & 0.43 & 0.73  &  0.73   & 0.54 & 0.69 &    1.29 & 0.56 & 0.74 &$-$0.07    &$-$0.07    \\\hline
             &AL1      & 0.58 & 0.48 &    0.60 & 0.59 &  0.56 &    0.55 & 0.57 & 0.49 &    0.71 & 0.61 & 0.63 &    0.18   &    0.18 \\
             &AL2      & 0.60 & 0.44 &   0.62  & 0.62 & 0.57  &  0.53   & 0.60 & 0.42 &    0.77 & 0.62 & 0.56 &  0.13     & 0.13    \\
             &AP1      & 0.57 & 0.49 &   0.66  & 0.50 &   0.50&  0.57   & 0.57 & 0.48 &    0.73 & 0.59 & 0.62 &  0.20     & 0.20    \\
             &AP2      & 0.58 & 0.48 &   0.60  & 0.59 &  0.56 &  0.55   & 0.57 & 0.49 &    0.71 & 0.61 & 0.63 &  0.18     & 0.18    \\
$\Omega^*$   &BD       & 0.79 & 0.83 &   0.84  & 0.39 & 0.69  & 0.75    & 0.51 & 0.67 &    1.32 & 0.55 & 0.72 &$-$0.07    &$-$0.07    \\\hline
\end{tabular}
\caption{Same as Table~\protect\ref{tab:charmparam}, for the bottom sector.}
\label{tab:bottomparam}
\end{table}
%

%\newpage

\end{document}